# Theory meets experiment: insights into structure and magnetic properties of Fe$_{1-x}$Ni$_x$B alloy

Gourab Bhaskar,[1] Zhen Zhang[2], Yaroslav Mudryk,[3] Sergey L. Bud'ko[2,3], Vladimir P. Antropov,[2,3*] Julia V. Zaikina[1*]

[1] Department of Chemistry, Iowa State University, Ames, Iowa 50011, United States
[2] Department of Physics and Astronomy, Iowa State University, Ames, Iowa 50011, United States
[3] Ames National Laboratory, US DOE, Iowa State University, Ames, Iowa 50011, United States

* Corresponding authors.
Email addresses: antropov@ameslab.gov, yzaikina@iastate.edu

**Abstract.** We studied the structural and magnetic properties of the solid solution Fe$_{1-x}$Ni$_x$B through theoretical and experimental approaches. Powder X-ray diffraction, X-ray Pair Distribution Function analysis, and energy dispersive X-ray spectroscopy reveal that the Fe$_{1-x}$Ni$_x$B solid solution crystallizes in the β-FeB structure type up to x = 0.6-0.7 and exhibits anisotropic unit cell volume contraction with increasing Ni concentration. Magnetic measurements showed a transition from ferromagnetism to paramagnetism around x = 0.7. For x = 0.5, the low (< 0.3 μ$_B$) magnetic moments suggest itinerant magnetism despite the relatively high Curie temperature (up to 225 K). Theoretical calculations indicated different types of magnetic orderings depending on the Fe/Ni atomic order, with the antiferromagnetic state being stable for ordered FeNiB$_2$, whereas the ground state is ferromagnetic for the disordered alloy. Calculations also predicted the coexistence of low- and high-spin states in Fe atoms around the composition with x=0.5, in line with the experimental evidence from $^{57}$Fe Mössbauer spectroscopy. The two magnetically distinct Fe sites for x = 0.3, 0.4, and 0.5 observed by $^{57}$Fe Mössbauer spectroscopy can also be interpreted as two magnetically different regions or clusters that could affect the critical behavior near a quantum magnetic transition based on a potential ferromagnetic quantum critical point identified computationally and experimentally near x=0.64. This work highlights the complex interplay between structure and magnetism in Fe$_{1-x}$Ni$_x$B alloys, suggesting areas for future research on quantum critical behavior.

**Introduction**

Boride alloys of iron with other 3$d$ metals demonstrate a variety of magnetic properties with rapidly changing character of magnetic interactions depending on the identity and concentration of 3$d$-metal (see [1,2] and more recent publications [3-10]). For instance, monoborides of 3$d$ metals with Mn and Fe are strong ferromagnets with large magnetic moments and the Curie temperatures ($T_C$). In contrast, monoborides of Cr, Co and Ni are para- or diamagnetic. No antiferromagnetism was previously observed for the 3$d$-metal monoborides. While monoborides with two 3$d$-metals typically form disordered alloys, the antiferromagnetic (AFM) state was recently predicted to be stable for ordered FeNiB$_2$ (Figure 1) [11]. FeNiB$_2$ was also predicted to belong to the class of the so-called Highly Responsive State (HRS) materials, which are very sensitive to external perturbations (pressure, magnetic fields, etc.). The HRS is found in proximity to magnetic instability or Quantum Critical Point (QCP), and one can expect that magnetic fluctuations are significant in HRS materials. Correspondingly, using static Density Functional Theory (DFT) calculations, the authors of [11] predicted strong AFM spin fluctuations, which can destroy the AFM state and produce superconducting instability.

Because gradual replacement of the transition metal in ferromagnetic monoborides MnB or FeB with Cr, Co, or Ni affords fine-tuning of the position of Fermi level, (Cr, Co, Ni)-doped (Mn, Fe)B disordered alloys provide a fruitful playground to study magnetic-to-non-magnetic transitions with an expected emergence of QCPs. The emergent QCP is often related to unusual chemical or magnetic local ordering and has recently been a subject of numerical, experimental, and analytical studies [12-15]. However, studies of criticality in disordered alloys of 3$d$-metal monoborides have not been reported. Among these systems, FeB is a strong ferromagnet [16], while NiB is either diamagnetic or weakly paramagnetic [17]. Therefore, we propose the (Fe$_{1-x}$Ni$_x$)B solid solution as a platform to study magnetic Quantum Phase Transitions (QPTs) by reducing the ferromagnetic ordering temperature $T_C$ down to zero and stabilizing a paramagnetic phase (PM) without magnetic order so that the QPT emerges when the ferromagnetic state transitions to the paramagnetic state.

The experimental study of magnetic QCPs is not a simple task. While substitution and pressure can effectively tune the system to the QCP, the accompanying variations in local atomic and structural inhomogeneities can strongly affect the critical behavior near the QCP, as homogeneous and disordered FM quantum critical transitions are very different [12, 15]. The different types of disorder, especially finite-sized magnetic clusters, can lead to the appearance of exotic QCPs. Thus, the structural characterization of the nature of the disorder near a magnetic QCP is essential to understanding the physics of quantum criticality in a given system.

From a theoretical perspective, the standard DFT methods for calculating electronic structure demonstrated remarkable success, correctly describing the concentration dependencies of magnetic moment and $T_C$ in many 3$d$-metal magnetic alloys [18]. For magnetic borides, not only experimental magnetization and $T_C$ were successfully confirmed via computations [6,9], but also such delicate properties as Spin Reorientation Transitions (SRTs) have been explained by the traditional band theoretical methods. For instance, the disordered substitutional alloy (Fe$_{1-x}$Co$_x$)$_2$B at x=0.3 exhibits three concentration-driven SRTs [19]. Such spin reorientation was rationalized by the filling of electronic bands with increasing electron concentration [20]. Another peculiar feature was that the anisotropy change was observed by varying the measurement temperature at a given composition [19-21]. Previous and recent experimental measurements of magnetic anisotropy constant as a function of dopant concentration, x, are also well reproduced by first-principles electronic structure analyses [20, 21].

An interesting feature of such disordered alloys is the possible appearance of QCPs in itinerant magnets [12]. While traditional theoretical models use Heisenberg or Ising models, they are not applicable to itinerant magnets. Itinerant magnetism appears because of collective interactions between atomic clusters and their surroundings. The itinerancy of the system in general and near its QCP cannot be directly identified experimentally, while it can be described using modern electronic structure calculations [18]. In turn, electronic structure calculations, while widely used, often fall short of providing definitive answers. The different choice of methods and generality of the exchange-correlation potentials create a large uncertainty. As such, the synergistic feedback loop between theory and experiments is crucial, enabling the successful computational description of different properties of a system before the prediction of a new property.

In this paper, we focus on iterative experimental and theoretical studies to analyze the structural and magnetic properties of the $Fe_{1-x}Ni_xB$ solid solution. We demonstrate how this multidisciplinary approach allows us to build a consistent description of the magnetism of a complex alloy. Before experimental studies of magnetic properties, we started with an analysis of the possible disorder and clustering in $(Fe_{1-x}Ni_x)B$ solid solution. We found that Ni and Fe distribution is disordered within the metal sublattice in this solid solution. Strong FM coupling was found in all alloys with x<0.6 with no traces of superconductivity at any studied Ni concentration. Consecutive theoretical electronic structure analysis demonstrated that observed chemical disorder is responsible for the stabilization of itinerant FM order in these borides. Calculations also revealed the possible coexistence of low- and high-spin states of Fe atoms. Analysis of $^{57}Fe$ Mössbauer spectroscopy data confirmed the low- and high-spin Fe states in the disordered $Fe_{1-x}Ni_xB$ solid solution, validating the itinerant mechanism of the onset of ferromagnetism.

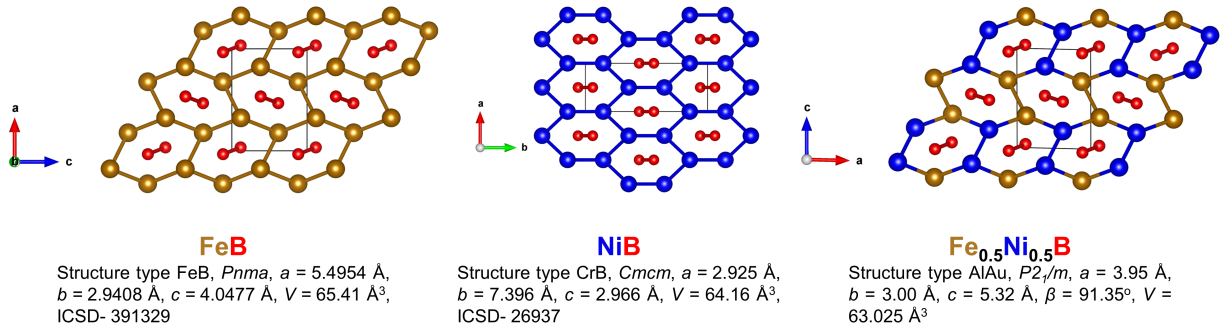

**FeB**
Structure type FeB, *Pnma*, $a$ = 5.4954 Å, $b$ = 2.9408 Å, $c$ = 4.0477 Å, $V$ = 65.41 Å$^3$, ICSD- 391329

**NiB**
Structure type CrB, *Cmcm*, $a$ = 2.925 Å, $b$ = 7.396 Å, $c$ = 2.966 Å, $V$ = 64.16 Å$^3$, ICSD- 26937

**Fe$_{0.5}$Ni$_{0.5}$B**
Structure type AlAu, $P2_1/m$, $a$ = 3.95 Å, $b$ = 3.00 Å, $c$ = 5.32 Å, $β$ = 91.35°, $V$ = 63.025 Å$^3$

Figure 1. Crystal structures of β-FeB [ICSD- 391329], NiB [ICSD- 26937] and calculated ordered model of Fe$_{0.5}$Ni$_{0.5}$B, for which superconductivity was theoretically predicted [11].

**Experimental Section.**

*Synthesis.* Starting materials for the syntheses were powders of iron Fe (AlfaAesar, 99.998%), nickel Ni (AlfaAesar, 99.996%), and boron B (AlfaAesar, amorphous & crystalline, 98%). All manipulations of the starting materials were performed in an argon-filled glovebox with $p(O_2) < 1$ ppm. Powders of the starting materials were weighed in the desired molar ratio Fe:Ni:B = (1-x):x:1.2, where x = 0, 0.3-0.7 (total mass 0.5 g) and loaded into polystyrene grinding vial sets with slip-on caps and polystyrene balls. A 20 mol. % excess of boron was used for the synthesis, while the stoichiometric molar ratio Fe:Ni:B = (1-x):x:1 without excess boron results in the impurities of other metal-rich binary borides, $Fe_2B$ and $Ni_2B$. Each grinding vial was sealed into double polypropylene bags under an argon atmosphere to prevent possible oxidation and brought out of the glovebox for the powder mixing/milling using a ball mill (SPEX 8000M MIXER/MILL). The starting materials were ball-milled for 1 hour to ensure homogeneous mixing. Inside the glovebox, ball-milled powders were loaded in a stainless steel die (inner diameter 10 mm) and cold-pressed to form pellets using a hydraulic press (Across) under ~900 MPa pressure. The pellets were arc-melted inside an argon-filled glovebox, flipped, and remelted multiple times to ensure homogeneity. Arc-melted ingots were sealed in niobium tubes and placed into a silica reactor equipped with a Swagelok safety check valve to maintain an $O_2$-free environment and prevent over-pressurization. The silica reactor was evacuated to the $4\times10^{-5}$ Bar and placed in resistance furnaces connected to temperature controllers. Samples were heated from room temperature to 1223 K in 8 hours, annealed at 1223 K for 7 days, and cooled down to room temperature by switching off the furnace. A pure sample of FeB was synthesized by arc-melting iron and boron powders in the Fe:B = 1:1.2 molar ratio without subsequent annealing. The synthesized compounds are stable in air and moisture for months.

Unit cell parameters of samples from the solid solution $Fe_{1-x}Ni_xB$ with a wide homogeneity range (x = 0.3-0.7) have been compared to the reported parameters of $Fe_{1-x}Ni_xB$ with the same Fe/Ni ratio in work by Gianoglio et al. [22]. These reported parameters have been extracted from figures published in [22] using the Web Plot Digitizer [23].

*Powder X-ray Diffraction (PXRD).* The purity of the synthesized sample was confirmed from X-ray powder diffraction using a Rigaku MiniFlex600 powder diffractometer with Cu Kα radiation ($\lambda$ = 1.54051Å) and Ni Kβ filter. Data were collected on a zero-background plate holder in the air at room temperature. Rietveld refinement of the PXRD data was performed using the GSAS II software package [24].

*Synchrotron X-ray Pair distribution function (X-ray PDF).* Total X-ray scattering data suitable for PDF analysis were collected at beamline 11-ID-B at Advanced Photon Source (APS), Argonne National Laboratory, for phase-pure samples of solid solution $Fe_{1-x}Ni_xB$ (x= 0, 0.3, 0.4, 0.5, 0.6). Powdered samples were filled in Kapton (polyimide) capillaries with a 1.1 mm outer diameter. High energy X-rays ($\lambda$ = 0.14320 Å) and a large amorphous silicon-based area detector (Perkin–Elmer) were used to collect data to high $Q$-values ($Q_{max}$ = 33 Å$^{-1}$) [25-27]. An empty capillary was used as a blank for background calibration. The data were corrected for background, and Fourier transformed to yield the PDF, G(r), using GSAS-II [24]. Structural models were refined against the PDF data within the PDFgui software package [28]. Since the solid solution $Fe_{1-x}Ni_xB$ crystallizes in the β-FeB structure type, the β-FeB structural model was used for the PDF refinement, but with unit cell parameters obtained upon Rietveld refinement of PXRD data. For the $Fe_{0.5}Ni_{0.5}B$ compound, two other structural models were also considered for the PDF refinement: (1) the theoretically predicted structural model of $FeNiB_2$ ($P2_1/m$) [11] (Figure 1), and

(2) the FeB structural model updated with unit cell parameters obtained based on the Rietveld refinement of PXRD data and metal sites with fixed 50%/50% site occupancy of Fe/Ni, while atomic coordinates for the Fe and Ni sites were not constrained to be the same. For every fitting of the PDF data, scale factor, quadratic atomic correlation factor (delta2), instrument parameters ($Q_{damp}$ and $Q_{broad}$), lattice parameters, symmetry-constrained atomic coordinates of Fe or Fe/Ni, and isotropic Atomic Displacement Parameters (ADP) of Fe or Fe/Ni and B were refined.

*Elemental Analysis*. An FEI Quanta-250 field emission scanning electron microscope (SEM) equipped with an Oxford X-Max 80 detector and an Oxford Aztec energy-dispersive x-ray spectroscopy (EDXS) analysis system was utilized for the elemental analysis of the samples. Powdered samples were sprinkled on aluminum pin-type stub holders with double-sided carbon tape for the measurement. The powders were analyzed using a 15 kV accelerating voltage with an accumulation time exceeding 60s.

*Magnetic measurements.* Magnetic properties of $Fe_{1-x}Ni_xB$ powders encapsulated in silica tubes (approx. weight 8-30 mg) were measured using Quantum Design SQUID MPMS XL-7 magnetometer between 2 and 380 K. High-temperature magnetic properties of the Fe-rich samples (x=0.3, 0.4) with a temperature of magnetic ordering above 300 K, were also verified using the VSM-Oven option of the Quantum Design PPMS using the pressed samples between 300 and 600 K. A minor mismatch between the magnetization data obtained by these two setups is likely related to the difference between the settle mode (sample stabilized at each temperature) that was used during MPMS measurements and the continuous sweep mode of VSM-Oven in PPMS, while correction for the oven stick signal as not applied.

*$^{57}$Fe Mössbauer spectroscopy.* Mössbauer spectroscopy measurements were performed using a SEE Co. conventional constant acceleration type spectrometer in transmission geometry with an $^{57}$Co (Rh) source, kept at room temperature. For the absorber, 40-60 mg of powders of the samples were mixed with a ZG-grade BN powder to ensure homogeneity. The absorber holder comprised two nested white Delrin cups. The absorber holder was locked in a thermal contact with a copper block with a temperature sensor and a heater and aligned with the $\gamma$—source and detector. The absorber was cooled to a desired temperature using a Janis model SHI-850-5 closed cycle refrigerator (with vibration damping). The driver velocity was calibrated by $\alpha$-Fe foil and all isomer shifts (IS) are quoted relative to the $\alpha$-Fe foil at room temperature. A commercial software package MossWinn [29] was used to analyze the Mössbauer spectra in this work.

*Theoretical Method.* Initial theoretical studies of $FeNiB_2$ [11] were performed only for the ordered structure. As the experiments clearly suggest a chemical disorder, here we employed the Green's Function-Based Linear Muffin-Tin Orbital method GF-LMTO [30] within DFT, which can treat such chemical disorders within the Coherent Potential Approximation (CPA) [30]. The exchange and correlation energy are treated within the spin-polarized Generalized Gradient Approximation (GGA) and parameterized by the Perdew-Burke-Ernzerhof (PBE) formula [31]. The structural type of the $FeNiB_2$ phase, with 2 formula units and a space group of $P2_1/m$ with parameters obtained in [11], was used for the ordered structure, while disorder at finite concentrations was also considered in the experimental β-FeB structure. The GF-LMTO was also used for the exchange coupling calculations. To estimate $T_C$ within CPA we used the KKR version [32,33] of spin-polarized RKKY approximation [34-36] implemented in a code developed in [30].

Large supercell calculations were performed using VASP [37,38], which employs the PBE exchange-correlation functional and the projector augmented wave method. The convergence thresholds were $10^{-5}$ eV for electronic self-consistency and 0.01 eV Å$^{-1}$ for structural optimization.

The Brillouin zone was sampled by the Monkhorst-Pack scheme with a k-point grid of $2\pi \times 0.033$ Å$^{-1}$ in the structural optimization and with a denser k-point grid of $2\pi \times 0.02$ Å$^{-1}$ in the static calculation. The 32-atom supercells of NiB (Ni$_{16}$B$_{16}$) with the NiB structure type and Fe substitutions were simulated. In such a supercell, single Fe atoms are sufficiently separated from each other. Three Fe concentrations were considered: one Fe atom (6.25%), two Fe atoms (12.5%), and eight Fe atoms (50%) per Ni$_{16}$B$_{16}$ supercell. Spin-polarized calculations were conducted for the Fe atoms with magnetic spin, while Ni and B atoms remain nonmagnetic after electronic self-consistency. Similar calculations were performed for 32-atom supercells of NiB and eight Fe (50%) per Ni$_{16}$B$_{16}$ supercell, but in the β-FeB structure type, consistent with experimental findings.

**Results and Discussion**

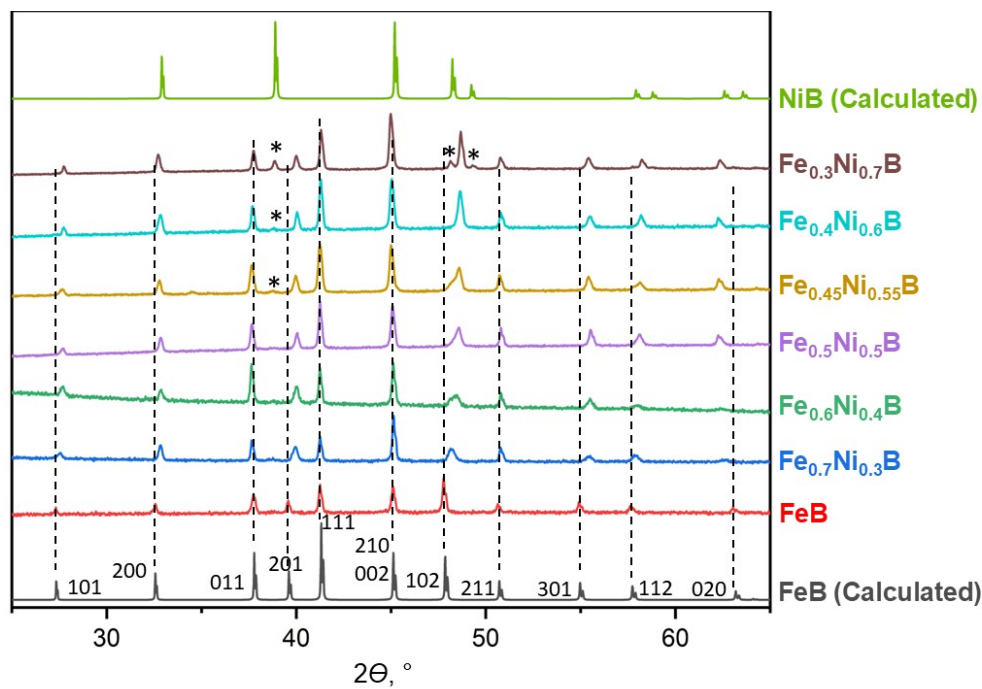

Figure 2. Powder X-ray diffraction patterns of the solid solution $Fe_{1-x}Ni_xB$ with variable Fe/Ni loading content. Calculated PXRD patterns of β-FeB (*Pnma*) [ICSD- 391329] and NiB (*Cmcm*) [ICSD- 26937] have been added for reference. Vertical dashed lines denote positions of the diffraction peaks corresponding to different (*hkl*) planes in the PXRD patterns. Peaks for NiB impurity are denoted with asterisk (*). The vertical axis corresponds to the relative diffraction intensity (arb. units).

## 1. *Synthesis and Powder X-ray Diffraction*

Recent computational predictions of superconductivity and spin fluctuations in the ordered $FeNiB_2$ ($Fe_{0.5}Ni_{0.5}B$) compound [11] (Figure 1) have led us to study the solid solution $Fe_{1-x}Ni_xB$ in a wide homogeneity range. FeB and NiB crystallize in two different orthorhombic structures: FeB in the β-FeB structure type (*Pnma*, $a$ = 5.4954 Å, $b$ = 2.9408 Å, $c$ = 4.0477 Å, $V$ = 65.41 Å$^3$, Z= 4, ICSD- 391329) and NiB in the CrB structure type (*Cmcm*, $a$ = 2.925 Å, $b$ = 7.396 Å, $c$ = 2.966 Å, $V$ = 64.16 Å$^3$, Z=4, ICSD- 26937) (Figure 1). Because FeB and NiB have different crystal structures, question about the structure and stability of the $Fe_{1-x}Ni_xB$ solid solution arises. Samples of the solid solution $Fe_{1-x}Ni_xB$ (x = 0.3-0.7) synthesized via arc-melting and subsequent annealing were found to crystallize in the FeB structure type, as evidenced by the similarity between powder X-Ray diffraction (PXRD) patterns of $Fe_{1-x}Ni_xB$ to that of FeB, even for Ni-rich samples (Figure 2).

Analysis of the PXRD data reveals a shift of the diffraction peaks with increasing Ni content, particularly corresponding to the (101), (200), (201) and (102) planes (Figure 2). This shift suggests the successful formation of the solid solution $Fe_{1-x}Ni_xB$ crystallizing in β-FeB structure type. Furthermore, the broadening/splitting of certain peaks, particularly (102) is noticeable, especially for x = 0.3-0.55. The broadening can be caused either by the Fe/Ni compositional fluctuations or by stacking faults and stochastic intergrowth of polymorphic slabs, alike the recently reported structure of α-FeB polymorph [39]. No additional Bragg peaks are observed, indicating an absence of superstructure from the ordered Fe/Ni arrangement. However, we note that similar X-ray scattering factors for Fe and Ni make an unambiguous determination of Fe/Ni ordering from X-ray scattering data impossible. From PXRD data we conclude that Fe and Ni atoms are randomly distributed in the $Fe_{1-x}Ni_xB$ lattice with no apparent superstructure, yet the PXRD data provides information about the average structure of a compound and the local Fe/Ni ordering in $Fe_{1-x}Ni_xB$ cannot be ruled out based on the PXRD data. The disorder in the Fe/Ni in $(Fe_{0.5}Ni_{0.5})B$ as opposed to the ordered $FeNiB_2$ phase studied computationally in [ref. 11] may result from the similar atomic radii and coordination environments of Fe and Ni, which provide no driving force for ordering, combined with high-temperature synthesis that favors entropy-driven disorder.

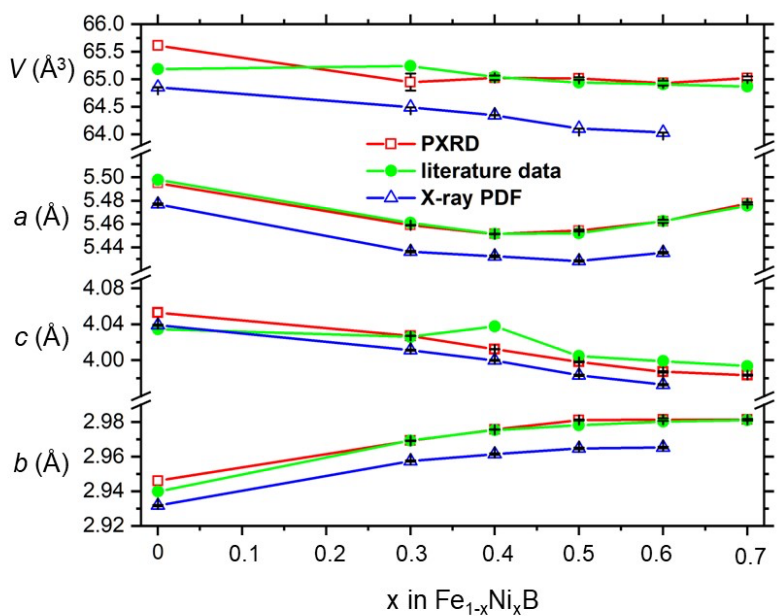

Figure 3. Variation of the unit cell parameters and unit cell volume with Ni content, x, in $Fe_{1-x}Ni_xB$ solid solution. The red squares and blue triangles represent the unit cell parameters/unit cell volume obtained in this work upon Rietveld refinement of PXRD data and X-ray PDF data, respectively, while filled green circles represent the unit cell parameters/unit cell volume obtained in the work by Gianoglio et al. [22]. The solid lines are drawn to guide the eye, and error bars (in black) are included data obtained in this work.

To understand the effect of Fe/Ni mixing, the unit cell parameters of $Fe_{1-x}Ni_xB$ samples have been determined via Rietveld refinement of PXRD data (Figure S1). Unit cell parameters obtained in this work for $Fe_{1-x}Ni_xB$ solid solution match relatively well with the values reported by Gianoglio et al. [22] (Figure 3). Owing to the substitution of bigger Fe with smaller Ni atoms in the $Fe_{1-x}Ni_xB$, an overall decrease in the unit cell volume is expected and observed experimentally, but only up to ~50% Ni substitution. For x > 0.5 (50%) in $Fe_{1-x}Ni_xB$, little-to-no change in the unit cell volume is observed within the 3 standard deviations, which, together with the increase in NiB impurity phase fraction from 6 wt.% for x = 0.6 to 20 wt.% for x = 0.7 (from Rietveld refinement) suggests that the substitution limit is between 60-70% (x = 0.6-0.7) in $Fe_{1-x}Ni_xB$. We also note that variation in the unit cell volume for different Ni loading content in $Fe_{1-x}Ni_xB$ does not follow Vegard's law, most likely due to variation in band filling. The change in unit cell volume is not linear, while the overall unit cell volume decreases with increasing Ni content. The unit cell contraction in $Fe_{1-x}Ni_xB$ is accomplished by a nearly linear decrease in the *c*-unit cell parameter. At the same time, the *b*-parameter gradually increases, and *a* parameter exhibits a non-monotonous change, first decreasing until x = 0.4 in $Fe_{1-x}Ni_xB$, followed by a linear increase for x>0.4. We conclude that the solid solution $Fe_{1-x}Ni_xB$ features anisotropic unit cell contraction upon substituting Ni for Fe. As it will be shown below by magnetic properties measurement and $^{57}Fe$ Mossbauer spectroscopy, the $Fe_{1-x}Ni_xB$ alloy exhibit bulk ferromagnetism at room temperature for x > 0.5 (*vide infra*). As such, the deviation in unit cell parameters/volume from the expected from Vegard law linear behavior can be attributed to the magnetic volume effect.

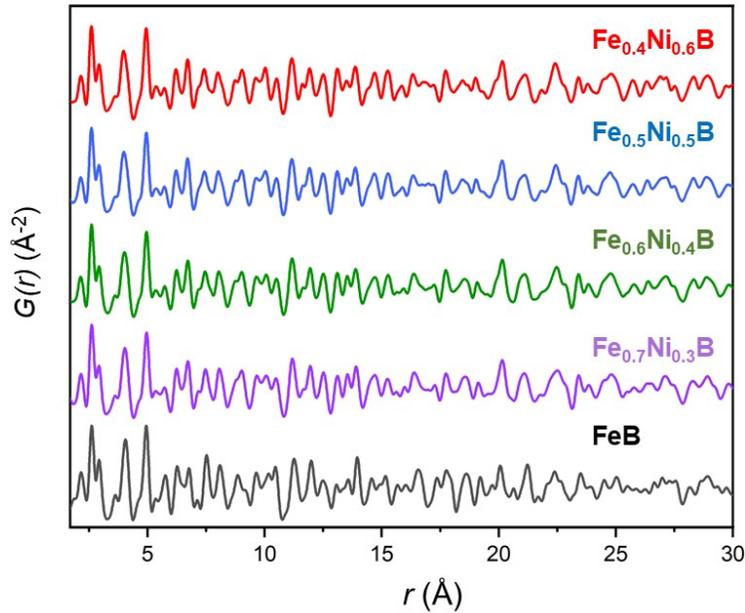

Figure 4. X-ray pair distribution function (PDF) data of the solid solution $Fe_{1-x}Ni_xB$ with variable Fe/Ni loading content in comparison to FeB.

## 2. *X-ray Pair Distribution Function.*

Analysis of X-ray pair distribution function (X-ray PDF) for the $Fe_{1-x}Ni_xB$ solid solution can provide insight into local ordering at the atomic scale. X-ray PDF is the Fourier transform of the total X-ray scattering data, which provides the probability of finding a pair of atoms separated by a distance, *r*. Therefore, the peaks in the PDF data represent various interatomic distances in the compound. As such, PDF is indispensable in obtaining information about any possible short-range ordering in crystalline or amorphous phases [40-42]. We note that the similar atomic/metallic radii for Fe and Ni imply similar bond distances in FeB and Ni, nevertheless, X-ray PDF data can be used to further confirm the formation of a solid solution.

A qualitative comparison of the PDF data of FeB and Fe-rich $Fe_{0.7}Ni_{0.3}B$ indicates slight changes in some of the peaks' relative intensities, suggesting minor changes in the coordination environment upon Fe/Ni mixing (Figure 4). However, minimal changes in the PDF data (primarily relative intensities) were observed between the samples with different Fe/Ni ratios in $Fe_{1-x}Ni_xB$. Since the solid solution $Fe_{1-x}Ni_xB$ crystallizes in the FeB structure type, the FeB structural model was used as the starting model for the PDF refinement, but with the unit cell parameters of $Fe_{1-x}Ni_xB$ obtained from PXRD data for variable values of x. Refinement of these structural models provided excellent fits of the PDF data for $Fe_{1-x}Ni_xB$ with $R_w$ values ranging from 0.14 to 0.12 (Figure S2-S6). Such an excellent fit suggests the formation of $Fe_{1-x}Ni_xB$ solid solution with β-FeB structure type, but a short-range Fe/Ni ordering cannot be ruled out based on these data. The slightly worse fit of X-ray PDF data for FeB ($R_w$ = 0.22, Figure S2) is attributed to the synthesis method used. FeB is known to form different polymorphs: in addition to thermodynamically stable β-FeB (FeB structure type), α-FeB was shown be a complex intergrowth of nanodomains with β-FeB and CrB structure types [42]. In this work, FeB was prepared via arc-melting of elemental Fe and B, while for the $Fe_{1-x}Ni_xB$ solid solution, a subsequent annealing of arc-melted ingots was employed. We hypothesize that the FeB that forms upon rapid cooling of arc-melted ingots contains stacking faults and CrB-like domains, while displaying the FeB structure type in bulk, thus explaining the mediocre fit of X-ray PDF data to the β-FeB structure type (Figure S2).

To further evaluate how X-ray total scattering data is affected by Fe/Ni distribution in $Fe_{1-x}Ni_xB$, we used three structural models (labeled as *A*, *B*, and *C*) for $Fe_{0.5}Ni_{0.5}B$ to fit X-ray PDF data: i) the FeB structural model (*Pnma*) with unit cell parameters obtained upon Rietveld refinement of the PXRD data of $Fe_{0.5}Ni_{0.5}B$ (model-*A*); ii) Ni atoms introduced into model-*A* as follows: Fe and Ni occupy the 4*c* site (x; ¼; z), where the x and z atomic coordinates for Fe and Ni are not constrained to be the same and Fe:Ni ratio was fixed to 0.5:0.5 (model-*B*) and iii) the theoretically predicted [11] structural model of $FeNiB_2$ (with monoclinic symmetry ($P2_1/m$) and ordered distribution of Fe and Ni (model-*C*) (Figure 1, Figure S7). Structural model-*A* provided an excellent fit of the PDF data with a low $R_w$ value of 0.125 and no correlations greater than 0.8. Fitting using model-*B* was attempted since PDF data is sensitive to interatomic distances, and the Ni atom is slightly smaller than the Fe atom, which may result in slightly shorter interatomic distances for Ni-Ni(B) than Fe-Fe(B). Even though structural model-*B* provided a somewhat better fit of the X-ray PDF data with a lower $R_w$ value of 0.102, four correlations between Fe and Ni coordinates and isotropic ADPs were detected, suggesting structural model-*A* should be preferred. Similarly, structural model-*C* also provided a slightly better fit of the X-ray PDF data with a low $R_w$ value of 0.118; however, in this case, five correlations between Fe and Ni coordinates and isotropic ADPs were observed, again suggesting that structural model-*A* is the most reasonable model (Figure S7). Unit cell parameters of $Fe_{1-x}Ni_xB$ samples refined from the X-ray PDF data compare well with cell parameters determined by Rietveld refinement of PXRD data and with the values reported by Gianoglio et al. [22] (Figure 3). All three datasets show similar trends and values of the unit cell parameters upon Fe/Ni mixing in the $Fe_{1-x}Ni_xB$, suggesting an anisotropic unit cell contraction. Hence, PXRD and PDF data analysis indicate that $Fe_{1-x}Ni_xB$ is a solid solution with β-FeB structure type Fe and Ni atoms to be randomly distributed in the $Fe_{1-x}Ni_xB$ with no apparent ordering. Comparing the fit quality of different models, we deem a formation of large Fe (Ni) clusters and long-range order of Fe and Ni atoms highly unlikely for all studied samples of the $Fe_{1-x}Ni_xB$ solid solution. However, we also note that the minor Fe/Ni clustering cannot be completely ruled out based only on the X-ray diffraction and total scattering data due to the similar X-ray scattering factors of Fe and Ni. Neutron diffraction or neutron PDF might not be insightful either, since the neutron scattering length for Fe and Ni are also similar (9.45 fm vs. 10.3 fm).

## 3. Composition from SEM-EDX

Scanning electron microscopy (SEM) images for two samples, $Fe_{0.6}Ni_{0.4}B$ and $Fe_{0.7}Ni_{0.3}B$, have been acquired to shed light on the Fe to Ni distribution in bulk (Figure 5). The homogeneous distribution of Fe and Ni atoms in bulk can be clearly seen from the false colored elemental maps, confirming the formation of $Fe_{1-x}Ni_xB$ solid solution, also evident from the PXRD and PDF data. Energy-dispersive X-ray spectroscopy (EDXS) verifies the presence of all three elements, Fe, Ni, and B, even though an accurate quantification of B cannot be achieved due to the limitations of the EDXS method with light elements. Signals for Nb (material of the tube used for synthesis) or any other heavy elements besides Ni and Fe were not detected in the samples. EDX was also utilized to quantify the Fe/Ni ratio in $Fe_{0.6}Ni_{0.4}B$ and $Fe_{0.7}Ni_{0.3}B$. The experimental ratios of 1.4(1) and 2.1(5), were determined by averaging the Fe/Ni ratio for 4 and 6 areas in each sample, respectively. These ratios agree well with the expected Fe/Ni ratio of 1.5 for $Fe_{0.6}Ni_{0.4}B$ and 2.33 for $Fe_{0.7}Ni_{0.3}B$. Dark spots present in the backscattered electron (BSE) images correspond to fractures that can be clearly seen upon comparing the BSE and secondary electron (SE) images (Figure S9).

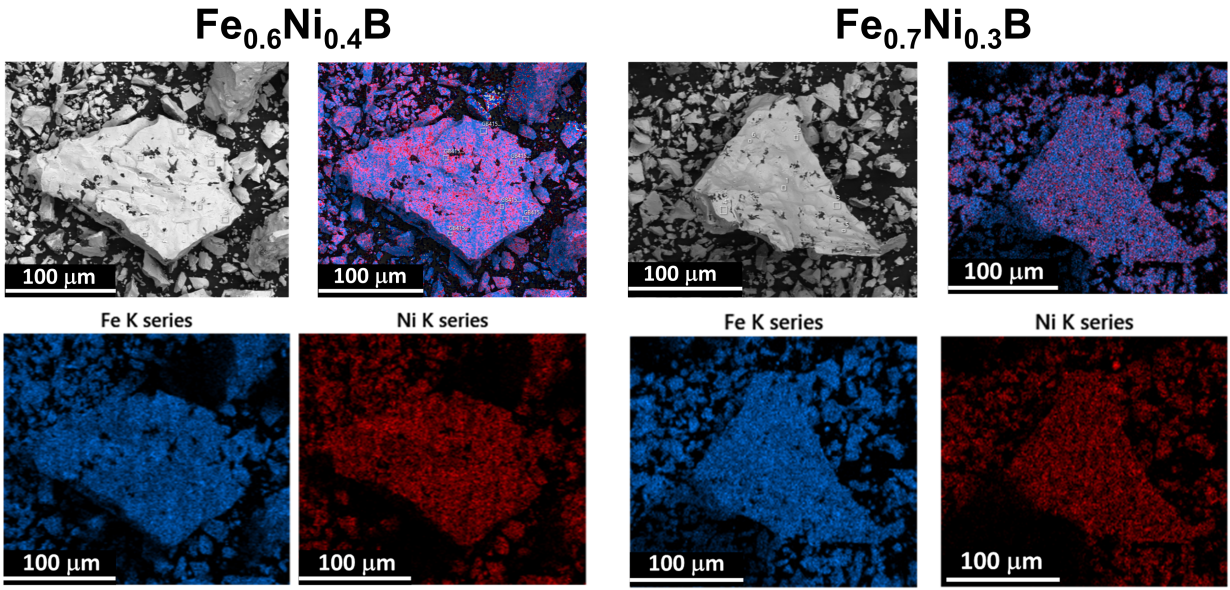

Figure 5. Back-scattered electron (BSE) images and false colored elemental maps illustrating the homogeneous distribution of Fe and Ni (blue and red, top), Fe (blue, bottom), and Ni (red, bottom) throughout the bulk samples of $Fe_{0.6}Ni_{0.4}B$ (left) and $Fe_{0.7}Ni_{0.3}B$ (right).

## 4. Magnetic Properties.

Magnetic properties, namely magnetization as a function of temperature, $M(T)$, and applied magnetic field, $M(H)$, as well as temperature- and frequency-dependent ac-susceptibility, $\chi_{ac}(T)$, were measured for $Fe_{1-x}Ni_xB$ powders with x = 0.3, 0.4, 0.5, 0.6, and 0.7. Bulk iron monoboride $\beta$-FeB samples obtained by arc-melting and subsequent annealing are ferromagnetic [16] with a Curie temperature of 590 K and magnetization $M = 1.1$ $\mu_B$/Fe atom at 4 K. Ni atoms bear no magnetic moment in NiB [17] and are reported as either a weak paramagnet or a diamagnet. Therefore, in the $Fe_{1-x}Ni_xB$ solid solution, magnetic behavior is expected to change from ferromagnetic to non-magnetic with an increase in x. The temperature dependencies of $M/H(T)$ were measured for all samples in the constant applied magnetic field of $\mu_0H = 0.1$ T (Figure S9) upon cooling and displayed the emergence of broad magnetic transitions for x = 0.3, 0.4, and 0.5. A steady decrease in magnetic ordering temperature and net magnetization is apparent when x increases in $Fe_{1-x}Ni_xB$. The $M/H(T)$ data of the Fe-rich samples (x ≤ 0.5) indicate paramagnetic-ferromagnetic long-range ordering, with magnetic ordering temperature, $T_C$, of $Fe_{0.7}Ni_{0.3}B$ reaching ~400 K (Figure S9a-b). In contrast, Ni-rich samples (x = 0.6 and 0.7) do not exhibit ferromagnetic ordering. The Fe-rich samples (x ≤ 0.5) show Curie-Weiss (CW) behavior in the paramagnetic region and the CW fitting of the data collected in the moderately high magnetic fields reveals paramagnetic Curie temperatures, $\theta_p$, of 386(2) K, 298(1), and 192(1) K, and paramagnetic moments of 1.9, 1.7, and 1.3 $\mu_B$/Fe for x=0.3, 0.4, and 0.5, respectively (Figure S10). The paramagnetic moment decline indicates reduced stability of Fe local moment with the increase in x(Ni content).

The magnetization as a function of the applied magnetic field, $M(H)$, measured at 2 K for all samples up to 7 T, shows the gradual increase in the magnetic moment per Fe atom with the increase in Fe content, reaching 0.74 $\mu_B$/Fe for $Fe_{0.7}Ni_{0.3}B$ (Figure 6). The $M(H)$ data at temperatures between 2 and 250 K for the $Fe_{0.5}Ni_{0.5}B$ compound present a conventional FM-like increase of magnetization with field, although full saturation is not reached even at 2 K (Figure S11).

The ac susceptibility measurements $\chi_{ac}$ were performed with the primary goal of assessing potential superconductivity and examining the ground state of compounds in the absence of a substantial external magnetic field (the drive amplitude of the ac magnetic field was set to $\mu_0 H_{ac}$ = 0.0005 T). The evolution of ac susceptibility with composition is shown in Figure 7. There is a broad maximum at ~90 K in $Fe_{0.6}Ni_{0.4}B$ and $Fe_{0.7}Ni_{0.3}B$, and a distinct downturn occurs below 20 K. As the Ni concentration increases, the broad peak at ~90 K transforms into a shoulder in the $Fe_{0.5}Ni_{0.5}B$ sample, while the low-temperature anomaly becomes more prominent as a clearly defined peak at 15 K, resembling an antiferromagnetic transition. The high-temperature shoulder practically disappears in $Fe_{0.4}Ni_{0.6}B$, while the temperature of the AFM-like anomaly shifts downward to 11 K. There are no magnetic anomalies in $Fe_{0.3}Ni_{0.7}B$ in the measured temperature interval from 2 to 40 K (Figure 7b). The plot of the imaginary part of measured ac susceptibility (Figure 7c) shows no evidence of long-range magnetic ordering in both $Fe_{0.3}Ni_{0.7}B$ and $Fe_{0.4}Ni_{0.6}B$ (Figures 7a and 7c). In agreement with Figure 6, the broad anomalies in $Fe_{0.5}Ni_{0.5}B$ (~225 K) and in $Fe_{0.6}Ni_{0.4}B$ (~320 K) are likely associated with ferromagnetic ordering (Figure 7a), and the onset of dissipative losses seen in the imaginary part of the ac susceptibility at these temperatures, indicates possible domain formation (Figure 7c). Thus, these temperatures can be assigned as Curie temperatures, although their values should be treated as approximate. We note, however, that $\chi''(T)$ does not follow a conventional FM behavior, and the presence of the broad peak indicates dissipative losses far below the observed $T_C$ that may be related to magnetic frustration. The magnetic ordering temperature of $Fe_{0.7}Ni_{0.3}B$, which, according to VSM-Oven data, is near 400 K, exceeds the measurement range of our MPMS SQUID system. The Curie temperatures determined from ac susceptibility data are in qualitative agreement with $M/H(T)$ data (Figure S10).

The remaining question is the nature of the low-temperature anomalies at ~15 and ~90 K. The peak at ~90 K is clearly observed in samples with ferromagnetic behavior (x=0.3, 0.4), and its weakening for x=0.5 and complete disappearance for x>0.5 suggests a correlation with the presence of FM ordering. The peak at 15 K is not seen in the imaginary part of ac susceptibility (Figure 7b-c), indicating that it is unrelated to ferromagnetism and local domain formation. Thus, the low-temperature anomalies may indicate the onset of AFM interactions. We note that these low-temperature transitions only occur in the near-zero-field range because the $M(H)$ data presented in Figure 6 and Figure S11 show conventional ferromagnetic behavior with no indication of metamagnetism. At the same time, the *M(H)* data also show no hysteresis and, correspondingly, no domain-wall pinning. The low-temperature anomaly could also mark the formation of a magnetic spin-glass system associated with the disorder in Fe/Ni mixed-occupied sites and competing AFM-FM interactions. To probe the 11 K anomaly in $Fe_{0.4}Ni_{0.6}B$, the ac susceptibility was measured at different frequencies of $\omega$ = 1, 10, 100, and 1,000 Hz (Figure 8). The peak exhibits weak but unambiguous frequency dependence typical for frustrated, potentially spin-glass, magnetism [43-45], namely the peak temperature, $T_f$, increases with *f*, from 10.8 K at 1 Hz to 11.3 K at 1,000 Hz, while the peak amplitude of real $\chi_{ac}$ decreases. This frequency dependence is weak and does not allow for a meaningful fitting analysis of $T_f(\omega)$, but its existence supports the presence of competing interactions. Notably, the ac susceptibility data (Figure 7) provide no evidence of potential superconductivity, which was theoretically predicted in the prior study [11]. Instead, as discussed above, our analysis shows the presence of complex low-temperature magnetic phenomena, except for the most Ni-rich composition $Fe_{0.3}Ni_{0.7}B$, which shows neither ferromagnetism nor superconductivity.

To summarize, FeB is a ferromagnet with a Curie temperature of 590 K and magnetic moment/Fe atoms = $1.1\mu_B$ [16] and NiB is diamagnet/weak paramagnet [13]. Ongoing from a Fe-rich to a Ni-rich phase in the $Fe_{1-x}Ni_xB$ solid solution, magnetic properties gradually change from ferromagnetic ordering to nonmagnetic behavior (Figures 6, 7, S9, S10). For the most Fe-rich composition, $Fe_{0.7}Ni_{0.3}B$, the magnetic moment/Fe atom ($0.74\ \mu_B$) and Curie temperature (~400 K) is lower than in FeB, consistent with dilution of Fe sublattice with non-magnetic Ni. We also observed Ni-rich phase $Fe_{0.3}Ni_{0.7}B$ to be paramagnetic; however, a slight increase in Fe content from 30% (x = 0.7) to 40% (x = 0.6) in $Fe_{1-x}Ni_xB$ resulted in magnetically frustrated behavior (Figures 7-8).

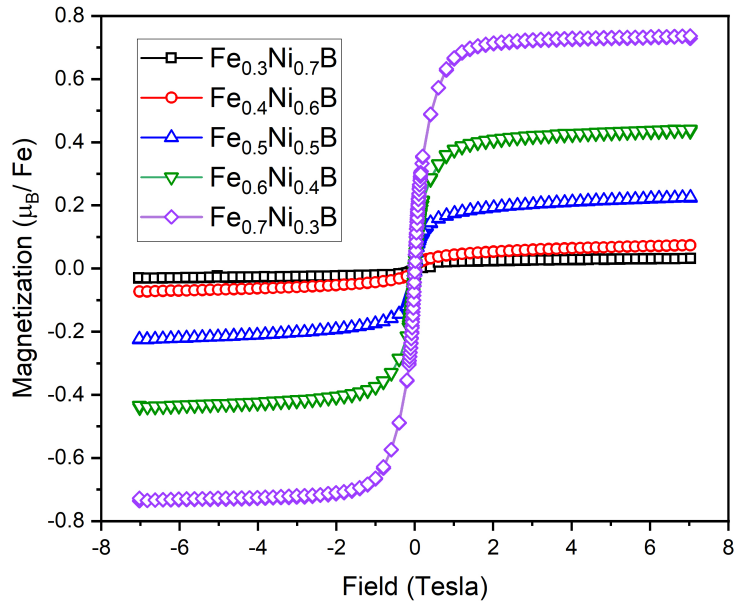

Figure 6. Magnetization of the $Fe_{0.3}Ni_{0.7}B$, $Fe_{0.4}Ni_{0.6}B$, $Fe_{0.5}Ni_{0.5}B$, $Fe_{0.6}Ni_{0.4}B$, and $Fe_{0.7}Ni_{0.3}B$ compounds as a function of the applied magnetic field measured at T = 2 K.

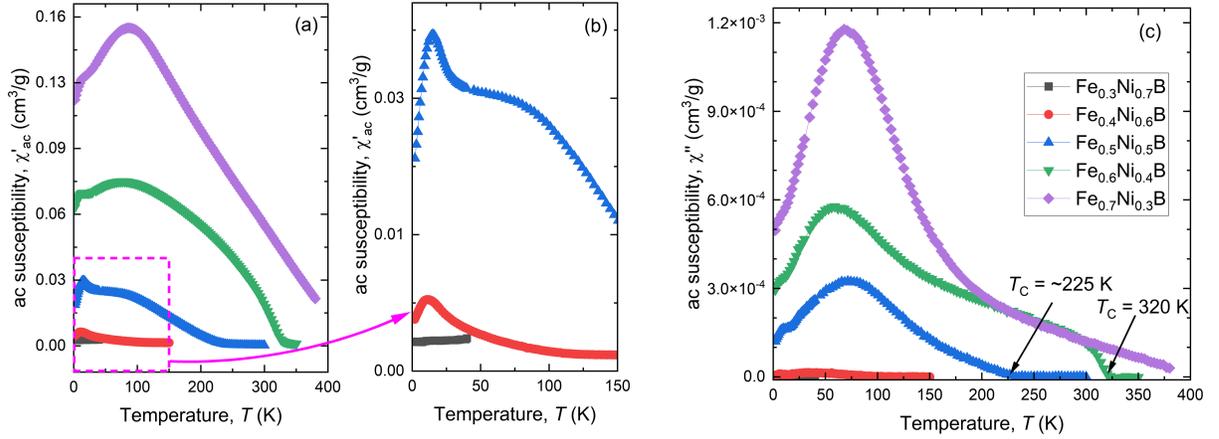

Figure 7. Temperature dependence of the ac magnetic susceptibility for the $Fe_{0.3}Ni_{0.7}B$, $Fe_{0.4}Ni_{0.6}B$, $Fe_{0.5}Ni_{0.5}B$, $Fe_{0.6}Ni_{0.4}B$, and $Fe_{0.7}Ni_{0.3}B$ compounds: a) real part, b) real part, insert from (a); and c) imaginary part.

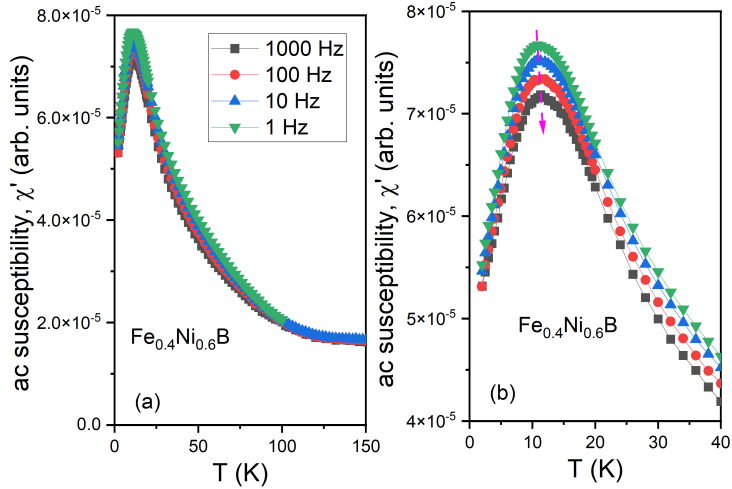

Figure 8. Frequency dependence of the real part of ac magnetic susceptibility in $Fe_{0.4}Ni_{0.6}B$.

## 5. Theoretical Calculations

Using input from the experiment, we performed electronic structure calculations, assuming a complete disorder between Fe and Ni atoms. To describe the disordered system, the CPA in GF-LMTO code was implemented [20, 30]. The PBE-optimized structure obtained in [11] was used with LDA self-consistency to reproduce the experimentally observed magnetic properties. These calculations revealed that such uniform disorder completely changes the character of magnetic interactions. While for the ordered $FeNiB_2$ system [11] (Figure 1) our calculations confirmed the previously found weakly stable AFM interaction between Fe atoms, the completely disordered $Fe_{0.5}Ni_{0.5}B$ alloy shows stable FM behavior. Moreover, the AFM state becomes less stable than the FM state with only 10% disorder in Fe/Ni sites within the ordered $FeNiB_2$ structure. The FM disordered state at x=0.5 appears to be by 25 meV more stable than the $FeNiB_2$ ordered state, suggesting an order-disorder transition with simultaneous magnetic AFM-FM phase transformation.

The computationally obtained magnetization (*M*) appears to align with experimental data. The theoretical point of ferromagnetism emergence is x = 0.64 and magnetization nearly linearly increases with further increase in Fe concentration. So, $M_{Fe}$ = 0.23 $\mu_B$ for at x=0.5 and $M_{Fe}$ = 0.69$\mu_B$ for x=0.3. For comparison, the magnetic moment in pure FeB (x=0) from our LDA CPA-TB-LMTO calculations is 1.13$\mu_B$.

To estimate $T_C$, we used the KKR version [31] of the spin-polarized RKKY approximation [32-34] implemented in a code developed in [20,30]. This approximation is better suited for relatively localized magnetic moments systems, while for more itinerant systems, the error can be above 20% (see discussion in [46]). Nevertheless, due to the error cancellation, the estimated $T_C$ describes concentration trends reasonably well [20,30]. Using the traditional mean field approximation $T_C$ = 2/3$J_0$, where $J_0$ is a zero moment of Heisenberg exchange couplings $J_0$ = $\sum J_{ij}$, we obtained 320 K for the disordered $Fe_{0.5}Ni_{0.5}B$, and 710 K for pure FeB. Though the experimental values for *T*c are consistently lower, by >15%, the obtained $T_C$ dependence over Fe content qualitatively predicts the behavior of $T_C$ in these alloys and the appearance of the long-range magnetic order. Single-site CPA describes satisfactory magnetism of 3*d*-metal based alloys, yet it does not include the effects of magnetic short-range order or local distortions for very diluted alloys. In our case, the appearance of magnetism at x=0.64 can indicate that magnetism has a pure itinerant nature and, to some extent, is similar to magnetism in the disordered Ni-Cu system, where it appears at around x = 0.5 [47].

Let us now discuss the magnetic behavior of these alloys at low concentrations of Fe atoms and consider the possible magnetic short-range order and local distortions around a Fe atom (impurity) by using large supercell VASP calculations. The single Fe atom impurity calculations revealed no magnetic instability in our 32-atom supercell (6.25% or one Fe atom in $Ni_{16}B_{16}$). This ultimately rejects any localized moment models, as the system represents an itinerant enhanced paramagnet with no local moments and strong metallic spin fluctuations. The exchange field addition from neighboring Ni atoms is too small for the Fe atom exchange field to generate atomic magnetic instability.

For the two Fe atom impurities (12.5% of Fe in $Ni_{16}B_{16}$), nearest neighbors (NN), second NN, third NN, and farthest possible neighbors within the used supercell were considered, and these models are shown in Figure 9(a-d), respectively. Only in the case of the second NN interaction (Figure 9b) was a stable FM ground state found for a pair of Fe atoms (0.41$\mu_B$). For all other cases of two Fe atom impurities, including the most uniform distribution (atoms are in different layers, Figure 9d), nonmagnetic ground states were revealed. Thus, our supercell calculations predict that already for concentrations around x = 0.875 in our system, some weakly interacting magnetic clusters can appear, although single-site CPA calculations predict the onset of magnetism only around x=0.64.

To assess a possible magnetic short-range order, we considered a supercell with a disordered distribution of Fe atoms at 50% concentration (8 Fe atoms in $Ni_{16}B_{16}$, Figure 9(e-f)). For the NiB structure, where Fe atoms exhibit two distinct magnetic states: Fe3, Fe4, Fe7, and Fe8 have smaller moments (0.11-0.15) $\mu_B$, while Fe1, Fe2, Fe5, and Fe6 have large moments (0.46-0.63) $\mu_B$. The NN pairs (Fe3, Fe4) and (Fe7, Fe8) exhibit antiparallel spin orientations within each pair (with overall zero magnetization), while the higher-spin Fe1, Fe2, Fe5, and Fe6 sites form pairs with parallel spin orientations. Thus, such random distributions lead to overall ferrimagnetic behavior with an average magnetic moment of about 0.27 $\mu_B$/Fe, which is close to our CPA results from above. For the FeB structure, Fe atoms also exhibit two distinct magnetic states: Fe4 and Fe5 have relatively larger moments (0.24-0.33) $\mu_B$, while the remaining six Fe atoms have close to zero moments. The magnetic behavior in the FeB structure is similar to that described above in the NiB structure, with somewhat smaller magnetic moments. Thus, at x=0.5, the randomly distributed Fe atoms in both structures form two distinct clusters: ferromagnetic clusters with large moments and antiferromagnetic clusters with small moments, coexisting within the system.

Overall, our results clearly indicate that the diluted $Fe_{1-x}Ni_xB$ solid solution with low Fe content is nonmagnetic, and the magnetic transition at x=0.64 appears because of the emerging itinerant interactions between Fe atoms. The magnetic moment does not form on isolated Fe atoms (Anderson criteria is not fulfilled). The overall magnetic moment forms only when there is an increase in the concentration of Fe atoms and when several other Fe atoms appear among the nearest neighbors for a given Fe atom. In such a case, the needed exchange field results in the development of local magnetic instability. Simultaneously, long-range order (in our case, primarily ferromagnetic) is developed among randomly distributed Fe atoms. Since the $Fe_{1-x}Ni_xB$ solid solution does not form for x > 0.7, we cannot establish the magnetic state of a single Fe impurity in the NiB structure type based on experimental results, and the discussion above is only a theoretical prediction. However, the overall good agreement between theory and experiment for x<0.6 in this study provides some confidence in this claim.

It is known that GGA satisfactorily describes geometry but tends to overestimate magnetic moments. Our GGA calculations confirmed the results obtained above with LDA method, but with all magnetic moments on Fe atoms being higher by about 0.2-0.3 $\mu_B$. The major difference is obtained for low Fe concentration cases (single atom impurities), where a Fe atom appears to be magnetic with a moment of 0.43 $\mu_B$. The effective interaction between Fe pairs appears to be predominantly ferromagnetic in GGA calculations. Thus, in GGA the calculations predict localized magnetism in these alloys.

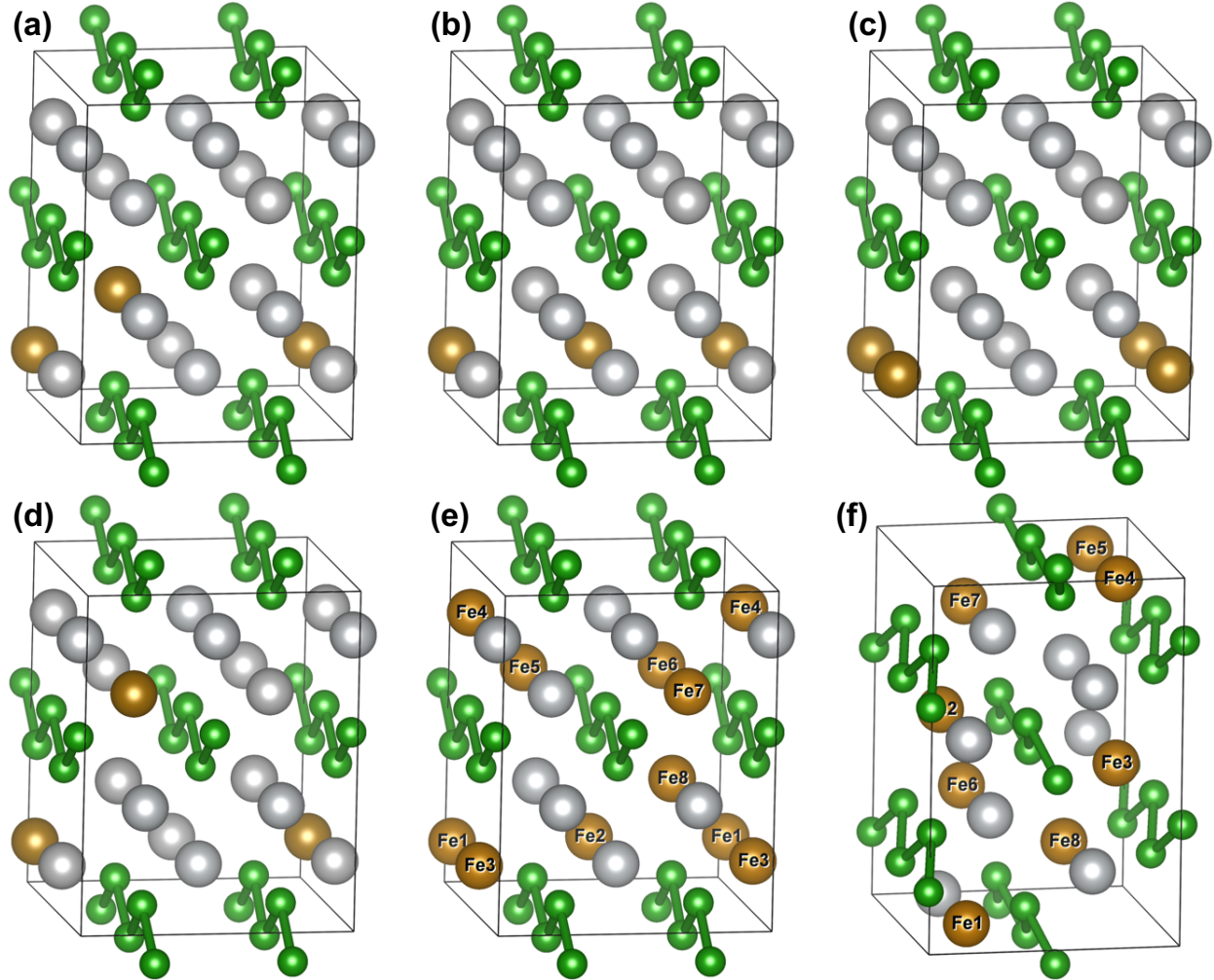

Figure 9. 32-atom supercells with 2 Fe impurities (12.5%, a-d) and 8 Fe impurities (50%, e,f). Fe positions are (a) Nearest neighbors (NN), (b) second NN, (c) third NN, (d) farthest neighbors within the used supercell with 2 Fe atom impurities (12.5). For the model with 50% of Fe atom impurities, Fe atoms are (e) disordered in the NiB structure type and (f) in the FeB structure type.

### 6. $^{57}$Fe Mössbauer spectroscopy.

To verify the computational prediction about the possible coexistence of two magnetic states in the $Fe_{1-x}Ni_xB$ alloys, we performed $^{57}$Fe Mossbauer spectroscopy studies. The experimental data (Figure 10) were fitted either with a doublet or with a combination of a doublet and a sextet. In the latter case, a reasonable fit was observed when the distribution of hyperfine fields on $^{57}$Fe sites was allowed. Data for the x = 0.6 samples between 5 K and 295 K were fitted with a doublet and no hyperfine field on $^{57}$Fe was observed. For samples with x = 0.5, 0.4, and 0.3 an additional sextet was required to fit the data in 5 – 75 K, 5 – 250 K, and 5 – 300 K, respectively, thus putting conservative lower boundaries for the ordering temperatures. Since for these three concentrations (x = 0.5, 0.4, 0.3), the percentage of the phase with finite hyperfine field on $^{57}$Fe atoms (e.g. with static magnetic moment on Fe atoms) changes with temperature in a close-to-linear fashion (Figure 11a). Somewhat more realistic, but still conservative magnetic ordering temperatures can be estimated from linear extrapolations. The estimated ordering temperatures are 130 K, 300 K, and 395 K for x=0.5, 0.4, and 0.3, respectively. The values of average hyperfine field for these three samples appear to be temperature and sample independent with $B_{hf} \sim 8$ T (Figure 11b), which roughly corresponds to the average magnetic moment of 0.5-0.6 $\mu_B$ per Fe (using a conversion factor of 15 T/$\mu_B$ for pure Fe, see [48] for a discussion of a proportionality coefficient between $B_{hf}$ and magnetic moment on Fe, and we will return to this evaluation below).

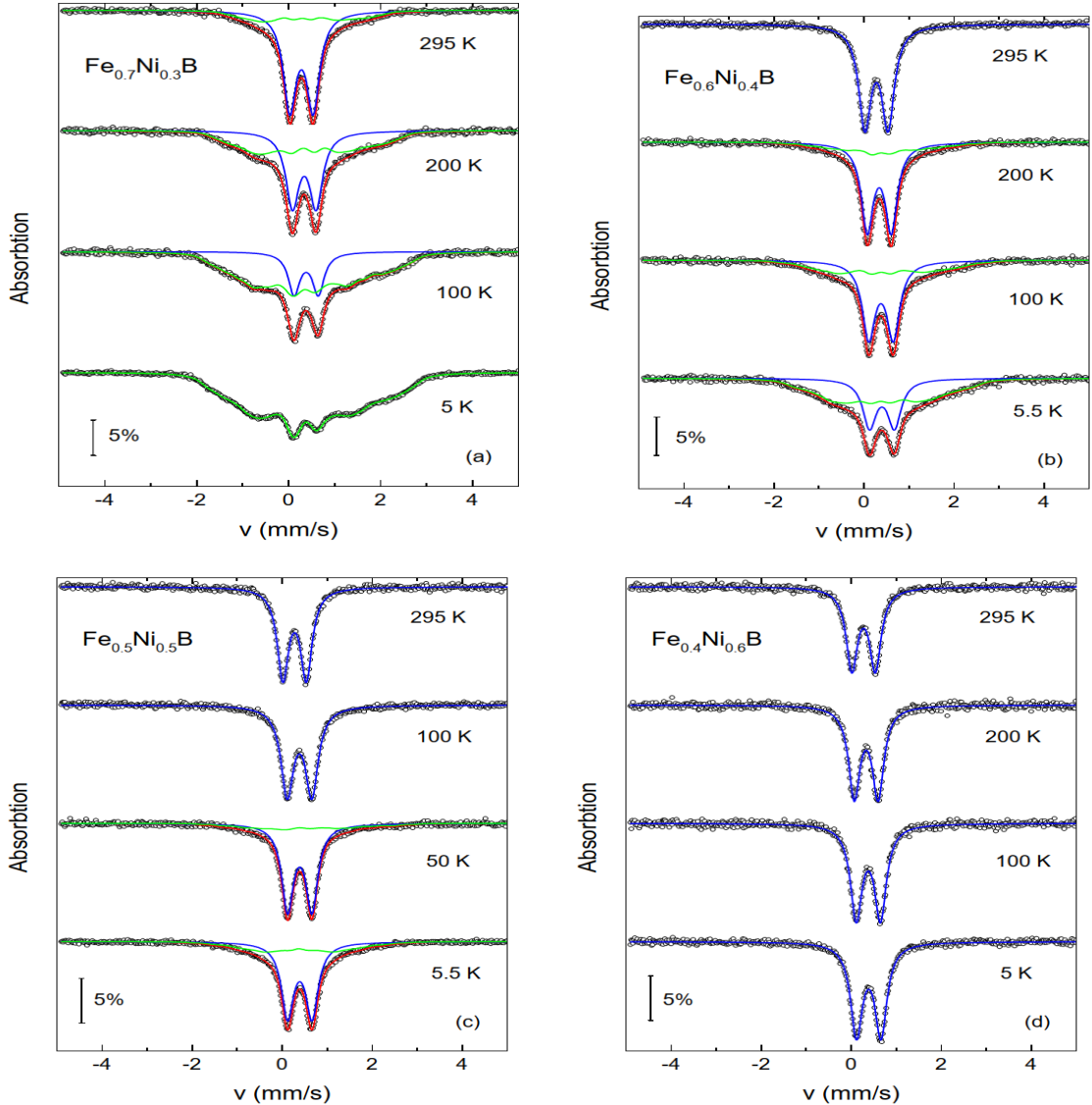

Figure 10. $^{57}$Fe Mössbauer spectra measured at different temperatures for Fe$_{1-x}$Ni$_x$B: (a) x = 0.3, (b) x = 0.4, (c) x = 0.5, (d) x = 0.6. Symbols are the experimental data, lines are the fits: blue – nonmagnetic doublet, green – magnetic sextet, red – doublet + sextet.

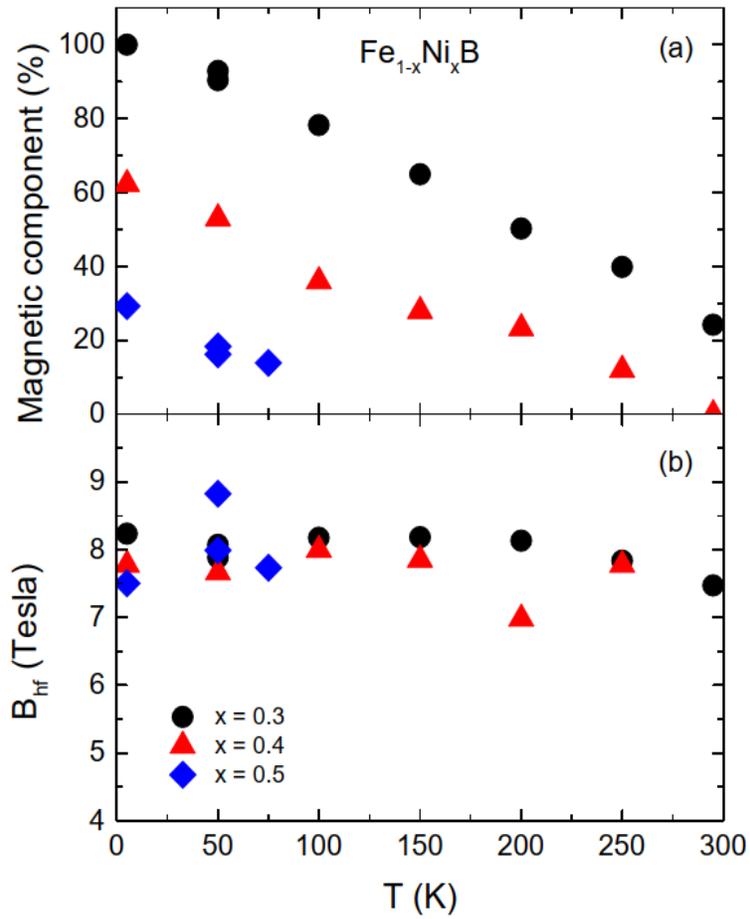

Figure 11. (a) Percentage of $^{57}$Fe atoms with finite hyperfine field (finite static moment) as a function of temperature for three $Fe_{1-x}Ni_xB$ samples with different Ni concentrations, x = 0.3, 0.4, 0.5. (b)Temperature dependence of the hyperfine field $B_{hf}$ obtained in $^{57}$Fe Mössbauer measurements for three $Fe_{1-x}Ni_xB$ samples with different Ni concentrations, x = 0.3, 0.4, 0.5.

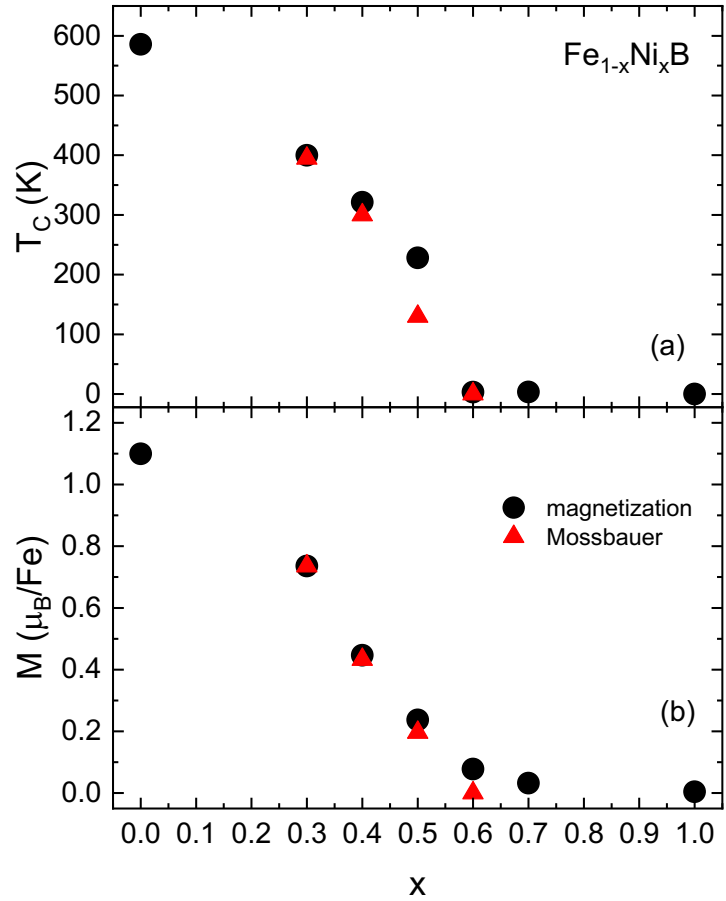

Figure 12. (a) Magnetic ordering temperature, $T_C$ (K), as obtained from the magnetization measurements (circles) and linear extrapolations of the $^{57}$Fe Mössbauer data from Figure 10 (triangles); (b) "average" magnetic moment on Fe atom as determined from 2 K $M(H)$ magnetization data (Figure 6) and as obtained from $^{57}$Fe Mössbauer data by multiplying of $B_{hf}$ by the percentage of magnetic phase and scaled to magnetization data for x = 0.3 sample.

Figure 12 presents a comparison of magnetic ordering temperature and "average" magnetic moment per Fe atom obtained from the magnetization measurements and $^{57}$Fe Mössbauer spectroscopy data. Except for the x=0.5 sample, the values of $T_C$ are consistent between two methods. The comparison of experimental magnetic moments from two different methods is more complex. $M(H)$ gives an average of the magnetic moment, whereas $^{57}$Fe Mössbauer spectroscopy yields a site-specific $B_{hf}$ (a proxy for the local magnetic moment). The value of $B_{hf}$ multiplied by the percentage of the magnetic component can serve as a measure of the average magnetic moment based on Mössbauer spectroscopy measurements. Indeed, the concentration dependence of an average magnetic moment (Figure 12a) is basically the same from the two measurements. Note that the scale factor for $B_{hf}$ in this figure is ~11.2 T/$\mu_B$, well within the ranges discussed in [48], yielding ~0.7 $\mu_B$/Fe moment on magnetic Fe sites, resulting in a $B_f$ of ~ 8 T (Figure 11). The computationally obtained moments for "high-spin" Fe sites are in the 0.46-0.63$\mu_B$ range (see above).

[57]Fe Mössbauer spectroscopy data agree reasonably well with the calculations. Both methods suggest two magnetically distinct Fe sites, where one site has no static magnetic moment associated with it. This result could also indicate two magnetically different regions or clusters that can affect the critical behavior near a quantum magnetic transition. The determination of the size of such clusters is a nontrivial task, both experimentally and computationally, and we leave this subject for future investigations.

It is noteworthy that for a conventional FM order, one would expect $B_{hf}(T)$ behavior to follow the saturation magnetization $M_{sat}(T)$ curve [49], which is not the case here. Moreover, in uniform materials with a single crystallographic Fe site, only a finite temperature region around magnetic transitions is characterized by the coexistence of doublets and sextet in Mössbauer spectra. This is, again, not the case here. These observations deserve further investigation and might point to the process of formation and growth of magnetic clusters in these compounds on cooling.

**Conclusion.** We performed theoretical and experimental studies of the structure and magnetic properties of $Fe_{1-x}Ni_xB$ alloys. Powder X-ray diffraction, X-ray Pair Distribution Function (X-ray PDF) analysis and elemental analysis using energy dispersive X-ray spectroscopy (EDXS) were used to study structural changes in $Fe_{1-x}Ni_xB$ polycrystalline samples with variation in x. This study confirms that the $Fe_{1-x}Ni_xB$ solid solution in the β-FeB structure type forms up to x ~0.6-0.7. The increase in Ni concentration leads to the anisotropic unit cell volume contraction, consistent with the smaller atomic size of Ni compared to Fe. X-ray PDF and EDXS point out the random Fe and Ni distribution within $Fe_{1-x}Ni_xB$ alloy, while Fe/Ni clustering cannot be ruled out given similar X-ray scattering factors for Fe and Ni.

Experimentally measured magnetization suggests that the magnetic properties of $Fe_{1-x}Ni_xB$ alloys gradually change from ferromagnetic to paramagnetic, as expected from the properties of the end-members: ferromagnetic FeB ($T_c$ = 590 K) and para/diamagnetic NiB. At low Fe concentrations, the low (below 0.3 $\mu_B$) magnetic moments in $Fe_{1-x}Ni_xB$ indicate the itinerant character of their magnetism. Despite such small moments, the measured Curie temperature in $Fe_{0.5}Ni_{0.5}B$ appears rather large (up to 225K). No superconductivity above 1.8 K was observed at any studied concentrations. Low-temperature frequency-dependent studies of ac magnetic susceptibility revealed a new AFM-like anomaly at around 11-15 K in $Fe_{0.5}Ni_{0.5}B$.

Our theoretical calculations revealed that $Fe_{1-x}Ni_xB$ at x=0.5 possess a different kind of magnetic ordering depending on the extent of chemical disorder between Fe and Ni atoms. While the ordered $FeNiB_2$ is weakly antiferromagnetic, a disordered alloy $Fe_{0.5}Ni_{0.5}B$ is strongly ferromagnetic. Calculations also predicted the coexistence of Fe atoms with low- and high-spin states around x=0.5 composition. The complementary experimental studies by [57]Fe Mössbauer spectroscopy suggest two magnetically distinct states of Fe atoms for x = 0.3, 0.4, 0.5. These two states can be interpreted as two magnetically different regions or clusters that can affect the critical behavior near a quantum magnetic transition. The possible ferromagnetic quantum critical point corresponding to the onset of ferromagnetism was obtained computationally at x=0.64 and is close to the experimental value. The detailed analysis of the behavior near this quantum critical point with the identification of quantum and rare phases will be the subject of further studies.

**Acknowledgments**
Z.Z., V.A. and J.V.Z. acknowledge financial support from the U.S. Department of Energy (DOE) Established Program to Stimulate Competitive Research (EPSCoR) Grant No. DE-SC0024284. Mössbauer spectroscopy (S.B.) and magnetic (Y.M) measurements were supported by the U.S. Department of Energy, Office of Basic Energy Sciences, Division of Materials Sciences and Engineering. Ames National Laboratory is operated for the U.S. Department of Energy by Iowa State University under Contract No. DE-AC02-07CH11358. This research used resources of the Advanced Photon Source, a U.S. Department of Energy (DOE) Office of Science user facility operated for the DOE Office of Science by Argonne National Laboratory under Contract No. DE-AC02-06CH11357.

**References**
[1] N. Lundquist, H. P. Myers, R. Westin, The paramagnetic properties of the monoborides of V, Cr, Mn, Fe, Co and Ni. *Philos. Mag.* **1962**, 7, 1187.
[2] M. C. Cadeville, E. Daniel, Sur la structure électronique de quelques borures d'éléments de transition. *J. Phys.* **1966**, 27, 449.
[3] S. Carenco, D. Portehault, C. Boissière, Mézailles, C. Sanchez, Nanoscaled Metal Borides and Phosphides: Recent Developments and Perspectives. *Chem. Rev.* **2013**, 113, 7981.
[4] Zhdanova, O. V.; Lyakhova, M. B.; Pastushenkov, Yu. G. Magnetic Properties and domain structure of FeB single crystals. *Met Sci Heat Treat*. **2013**, *55*, 68–72.
[5] Z. Pu, T. Liu, G. Zhang, X. Liu, Marc. A. Gauthier, Z. Chen, S. Sun. Nanostructured Metal Borides for Energy-Related Electrocatalysis: Recent Progress, Challenges, and Perspectives. *Small Methods*, **2021**, 5, 2100699.
[6] P. Mohn, D.G. Pettifor The calculated electronic and structural properties of the transition-metal monoborides. *J. Phys. C: Solid State Phys*. **1988**, 21, 2829–2839.
[7] Y. Bourourou, L. Beldi, B. Bentria, A. Gueddouh, B. Bouhafs, Structure and magnetic properties of the 3d transition-metal mono-borides TM–B (TM=Mn, Fe, Co) under pressures. *J. Magn. Magn. Mater.* **2014**, 365, 23.
[8] C. Romero-Muñiz, J. Yan Law, L. M. Moreno-Ramírez, Á. Díaz-García, Vi. Franco. Using a computationally driven screening to enhance magnetocaloric effect of metal monoborides, *J. Phys. Energy,* **2023**, 5 02402.
[9] S. Ma, K. Bao, Q. Tao, P. Zhu, T. Ma, B. Liu, Y. Liu, T. Cui. Manganese mono-boride, an inexpensive room temperature ferromagnetic hard material, *Scientific Reports* **2017**, 7, 43759.
[10] H. Kim, D. R. Trinkle. Mechanical properties and phase stability of monoborides using density functional theory calculations. *Phys. Rev. Mater.* **2017**, 1, 013601.
[11] R. Wang, Y. Sun, V. Antropov, Z. Lin, C.Z. Wang, K.M. Ho. Theoretical prediction of a highly responsive material: Spin fluctuations and superconductivity in FeNiB$_2$ system. *Appl. Phys. Lett*. **2019**, *115*, 182601.
[12] M. Brando, D. Belitz, F. M. Grosche, and T. R. Kirkpatrick, Metallic quantum ferromagnets. *Rev. Mod. Phys*. **2016**, 88, 025006.
[13] H. Liu, E. Huffman, S. Chandrasekharan, R. K. Kaul. Quantum Criticality of Antiferromagnetism and Superconductivity with Relativity. *Phys. Rev. Lett.* **2022**, 128, 117202.
[14] M. Vojta, Y. Zhang, S. Sachdev. Competing orders and quantum criticality in doped antiferromagnets. *Phys. Rev. B* **2000**, 62, 6721.
[15] M. Vojta. Frustration and quantum criticality. *Rep. Prog. Phys*. **2018**, 81, 064501.

[16] Rades, S.; Kraemer, S.; Seshadri, R.; Albert, B. Size and Crystallinity Dependence of Magnetism in Nanoscale Iron Boride, α-FeB. *Chem. Mater*. **2014**, *26*, 1549–1552.

[17] Mutlu, R. H.; Aydinuraz, A. Effect of particle size on the magnetic properties of NiB. *J. Magn. Magn. Mater*. **1987**, *68*, 328-330.

[18] H. Akai, P. Dederichs, J. Kanamori. Magnetic properties of Ni- and Co-alloys calculated by KKR-CPA-LSD method. *J. de Physique Coll*. **1988**, 49 (C8), 23-24.

[19] Iga, A. Magnetocrystalline Anisotropy in $(Fe_{1-x}Co_x)_2B$ System. *Jpn. J. Appl. Phys*. **1970**, *9*, 415.

[20] Belashchenko, K. D.; Ke, L.; Däne, M.; Benedict, L. X.; Lamichhane, T. N.; Tarfour, V.; Jesche, A.; Bud'ko, S. L.; Canfield, P. C.; Antropov, V. P. Origin of the spin reorientation transitions in $(Fe_{1-x}Co_x)_2B$ alloys. *Appl. Phys. Lett*. **2015**, *106*, 062408.

[21] Lamichhane, T. N.; Palasyuk, O.; Antropov, V. P.; Zhuravlev, I. A.; Belashchenko, K. D.; Nlebedim, I. C.; Dennis, K. W.; Jesche, A.; Kramer, M. J.; Bud'ko, S. L.; McCallum, R. W.; Canfield, P. C.; Taufour,V. Reinvestigation of the intrinsic magnetic properties of $(Fe_{1-x}Co_x)_2B$ alloys and crystallization behavior of ribbons. *J. of Magn. and Magn. Mat*. **2020**, *513*, 167214.

[22] Gianoglio, C.; Badini, C. Distribution equilibria of iron and nickel in two-phase fields of the Fe-Ni-B system. *J. Mater. Sci*. **1986**, *21*, 4331.

[23] https://automeris.io/WebPlotDigitizer

[24] Toby, B. H.; Von Dreele, R. B. GSAS-II: the genesis of a modern open-source all purpose crystallography software package. *J. Appl. Cryst*. **2013**, *46*, 544–549.

[25] Chupas, P. J.; Qiu, X.; Hanson, J. C.; Lee, P. L.; Grey, C. P.; Billinge, S. J. L. Rapid-acquisition pair distribution function (RA-PDF) analysis. *J. Appl. Cryst*. **2003**, *36*, 1342-1347.

[26] Chupas, P. J.; Chapman, K. W.; Lee, P. L. Applications of an amorphous silicon-based area detector for high-resolution, high-sensitivity and fast time-resolved pair distribution function measurements. *J. Appl. Cryst*. **2007**, *40*, 463-470.

[27] Toby, B. H.; Madden, T. J.; Suchomel, M. R.; Baldwin, J. D.; Von Dreele, R. B. A scanning CCD detector for powder diffraction measurements. *J. Appl. Cryst*. **2013**, *46*, 1058-1063.

[28] Farrow, C. L.; Juhas, P.; Liu, J. W.; Bryndin, D.; Božin, E. S.; Bloch, J.; Proffen, Th.; Billinge, S. J. L. PDFfit2 and PDFgui: computer programs for studying nanostructure in crystals. *J. Phys.: Condens. Matter*. **2007**, *19*, 335219-335225.

[29] Klencsár Z (2023) MossWinn 4.0 Pre. http://www.mosswinn.com/

[30] Pujari, B. S.; Larson, P.; Antropov, V. P.; Belashchenko, K. D. Ab Initio Construction of Magnetic Phase Diagrams in Alloys: The Case of $Fe_{1-x}Mn_xPt$. *Phys. Rev. Lett*. **2015**, *115*, 057203.

[31] J.P. Perdew, K. Burke, K. and M. Ernzerhof, Generalized Gradient Approximation Made Simple. *Phys. Rev. Lett.* **1996**, 77, 3865-3868.

[32] A.I. Liechtenstein, M.I. Katsnelson, V.A. Gubanov, Local spin excitations and Curie temperature of iron. *Solid State Comm.* **1985**, 54, 327 (1985)

[33] A.Z. Menshikov, V.P. Antropov, G.P. Gasnikova, Yu.A. Dorofeyev, V.A. Kazantsev, Magnetic phase diagram of ordered $Fe_{1-x}Mn_xPt$ alloys. *J. of Magn. Magn. Mater.* **1987**, 65, 159.

[34] S. H. Liu, Quasispin model of itinerant magnetism: High-temperature theory. *Phys. Rev. B* **1977**, 15, 4281.

[35] R. E. Prange and V. Korenman, Local-band theory of itinerant ferromagnetism. IV. Equivalent Heisenberg model. *Phys. Rev. B* **1979**, 19, 4691.

[36] J. F. Cooke, Role of electron-electron interactions in the RKKY theory of magnetism. *J. Appl. Phys*. **1979**, 50, 1782.

[37] G. Kresse and J. Furthmüller, Efficiency of ab-initio total energy calculations for metals and semiconductors using a plane-wave basis set, *Comput. Mater. Sci.* **1996**, 6, 15-50.

[38] G. Kresse and J. Furthmüller, Efficient iterative schemes for ab initio total-energy calculations using a plane-wave basis set. *Phys. Rev. B* **1996**, 54, 11169-11186.

[39] F. Igoa Saldaña, E. Defoy, D. Janisch, G. Rousse, P.-O. Autran, A. Ghoridi, A. Séné, M. Baron, L. Suescun, Y. Le Godec, D. Portehault. *Inorg. Chem.* **2023**, 62, 5, 2073–2082.

[40] Proffen, T.; Page, K. L.; McLain, S. E.; Clausen, B.; Darling, T. W.; TenCate, J. A.; Lee, S.-Y.; Ustundag, E. Atomic pair distribution function analysis of materials containing crystalline and amorphous phases. *Z. Kristallogr*. **2005**, *220*, 1002–1008.

[41] Billinge, S. J. L. The Rise of the X-ray Atomic Pair Distribution Function Method: a Series of Fortunate Events. *Philos. Trans. R. Soc. A*. **2019**, *377*, 20180413.

[42] Keen, D. A. Total Scattering and the Pair Distribution Function in Crystallography. *Crystallogr. Rev*. **2020**, *26*, 143−201.

[43] Anand, V. K.; Adroja, D. T.; Hillier, A. D. Ferromagnetic cluster spin-glass behavior in PrRhSn$_3$. *Phys. Rev. B*. **2012**, *85*, 014418.

[44] Scholz, T.; Dronskowski, R. Spin-glass behavior of Sn$_{0.9}$Fe$_{3.1}$N: An experimental and quantum-theoretical study. *AIP Adv*. **2016**, *6*, 55107.

[45] Bhaskar, G.; Gvozdetskyi, V.; Batuk, M.; Wiaderek, K.; Carnahan, S. L.; Sun, Y.; Wang, R.; Wu, X.; Zhang, C.; Ribeiro, R. A.; Bud'ko, S. L.; Canfield, P. C.; Rossini, A. J.; Huang, W.; Wang, C. Z.; Ho, K. M.; Hadermann, J.; Zaikina, J. V. Topochemical deintercalation of Li from layered LiNiB: toward 2D *M*Bene. *J. Am. Chem. Soc*. **2021**, *143*, 4213–4223.

[46] V. P. Antropov, The exchange coupling and spin waves in metallic magnets: removal of the long-wave approximation, *J. of Magn. Magn. Mater.* **2003**, 262 (2), L192-L197.

[47] S. A. Ahern, M. J. C. Martin, W. Sucksmith, The spontaneous magnetization of nickel + copper alloys. *Proc. Roy. Soc.* (London) **1958**, 248, 145.

[48] S.M. Dubiel, Relationship between the magnetic hyperfine field and the magnetic moment, *J. of Alloys and Comp.* **2009**, 488, 18.

[49] Nagle, D. E., Frauenfelder, H., Taylor, R. D., Cochran, D. R. F. and Matthias, B. T. Temperature dependence of the internal fields in ferromagnets. *Phys. Rev. Lett.* **1960**, *5*, 364.

# Theory meets experiment: insights into structure and magnetic properties of $Fe_{1-x}Ni_xB$ alloy

Gourab Bhaskar,[1] Zhen Zhang[2], Yaroslav Mudryk,[3] Sergei L. Bud'ko[2,3], Vladimir P. Antropov,[2,3*] Julia V. Zaikina[1*]

[1] Department of Chemistry, Iowa State University, Ames, Iowa 50011, United States
[2] Department of Physics and Astronomy, Iowa State University, Ames, Iowa 50011, United States
[3] Ames National Laboratory, US DOE, Iowa State University, Ames, Iowa 50011, United States

*Supporting Information*

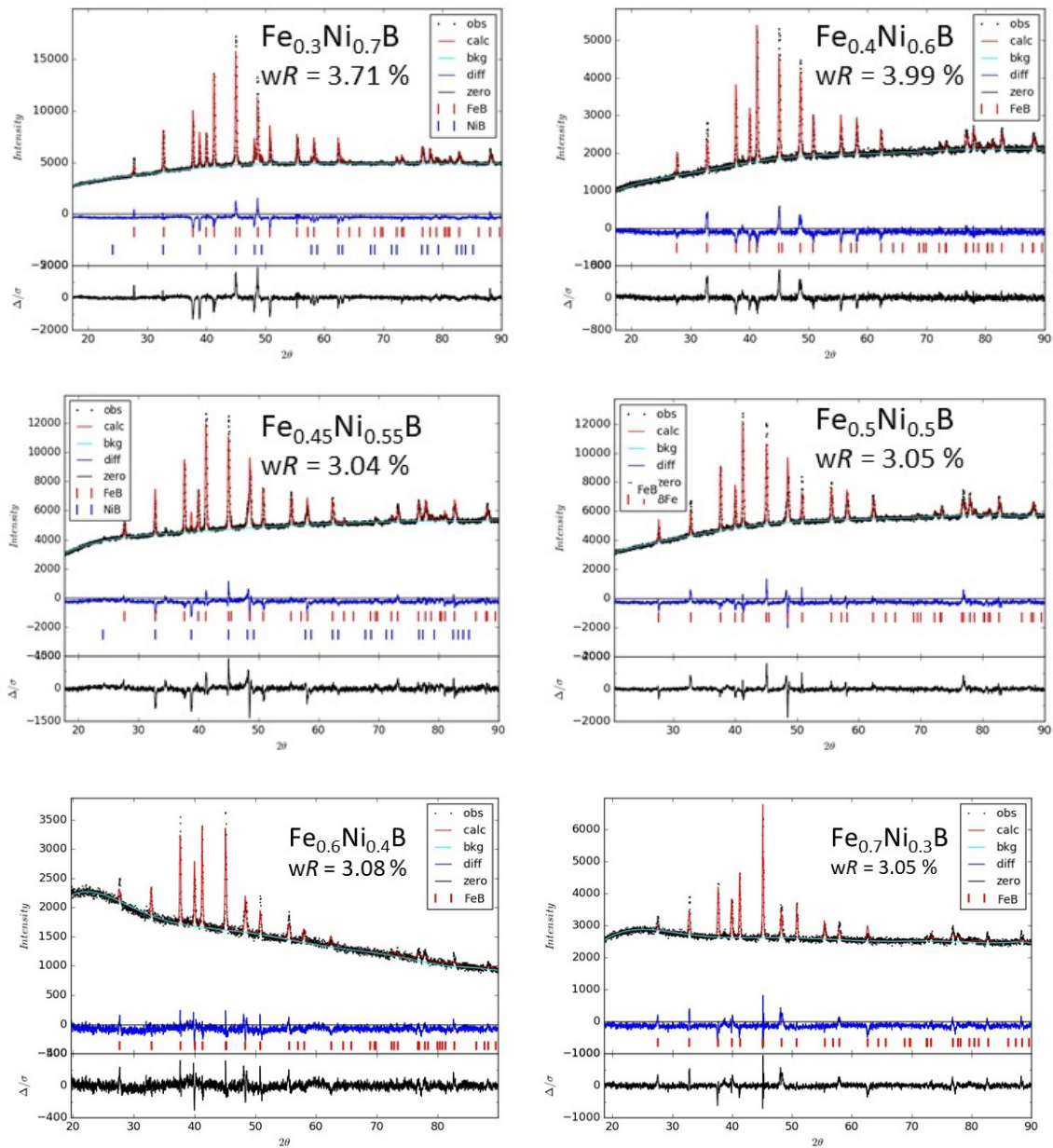

**Figure S1.** Rietveld refinement of laboratory PXRD data for the solid solution $Fe_{1-x}Ni_xB$ with variable Fe/Ni loading content ($x$). The structure models of FeB (*Pnma*) [ICSD- 391329] and NiB (*Cmcm*) [ICSD- 26937] were used for the Rietveld fit. The increased background is primarily due to fluorescence from Fe when exposed to CuKα X-ray radiation. The intensity mismatch is most likely due to the stacking faults and CrB-like domains within the FeB structure type in bulk, as reported for FeB polymorphs [F. Igoa Saldaña, E. Defoy, D. Janisch, G. Rousse, P.-O. Autran, A. Ghoridi, A. Séné, M. Baron, L. Suescun, Y. Le Godec, D. Portehault. *Inorg. Chem.* **2023**, 62, 5, 2073–2082].

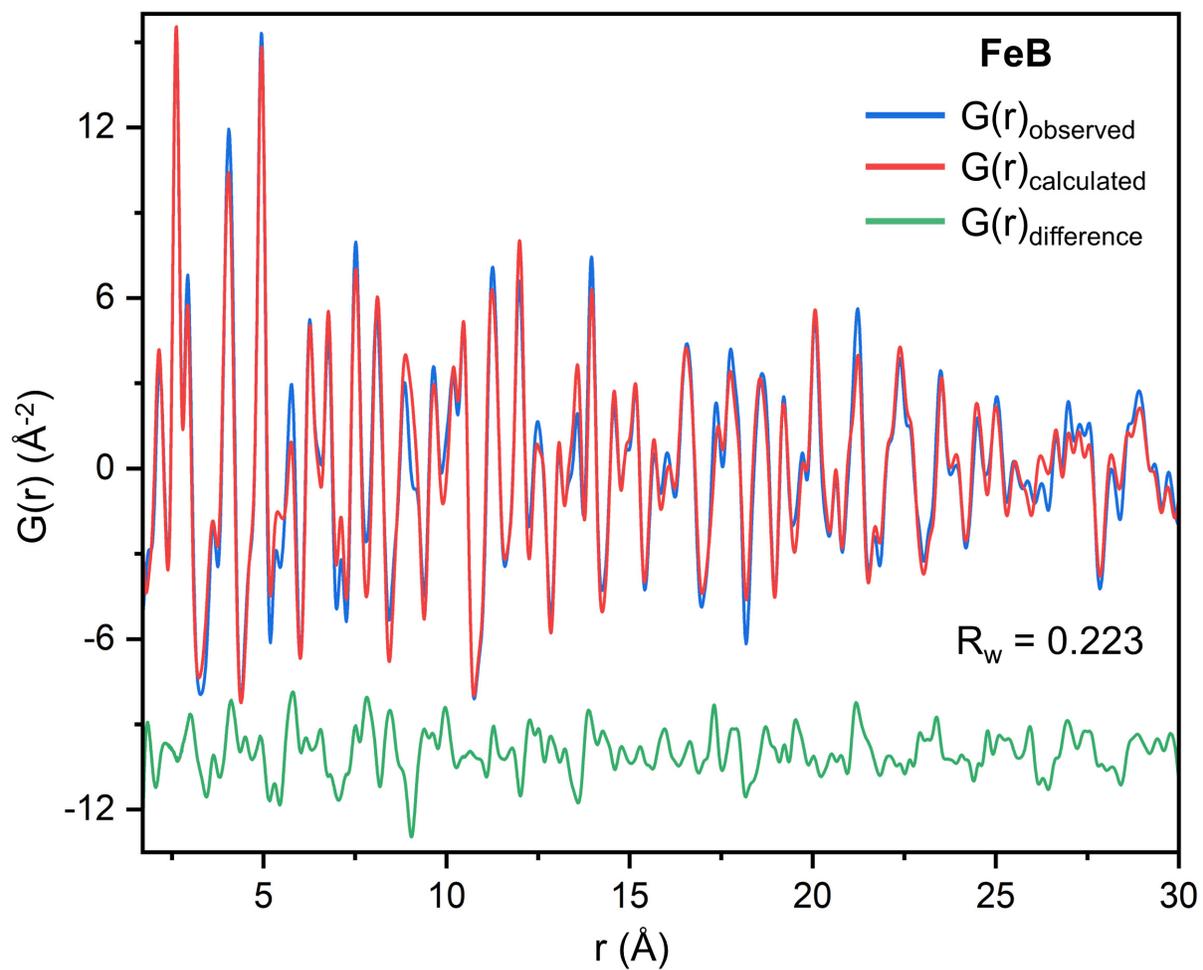

**Figure S2**. Fitting of the X-ray pair distribution function of FeB, $R_w$=0.223. For the refinement, FeB structural model with unit cell parameters obtained upon Rietveld refinement of PXRD data was utilized.

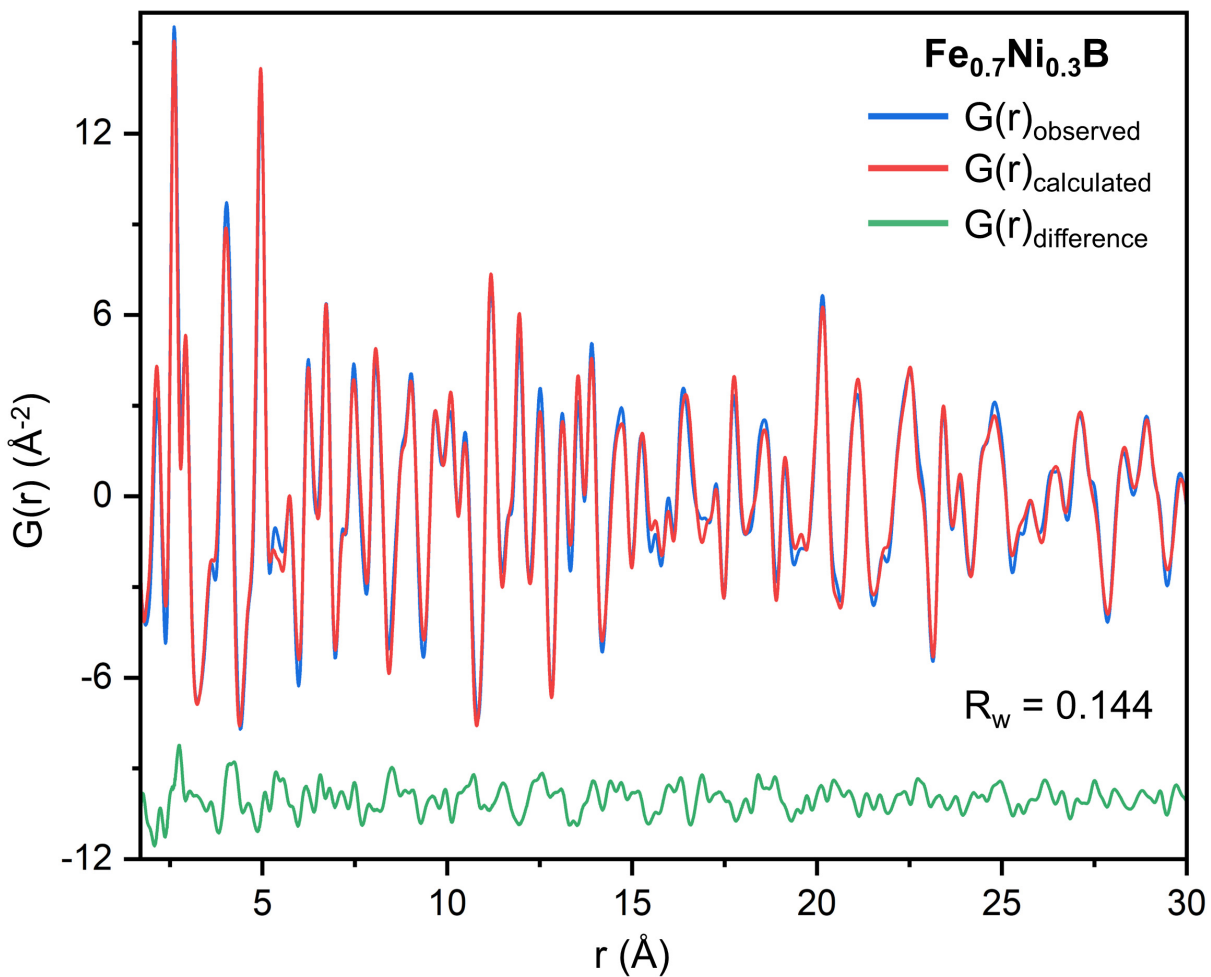

**Figure S3.** Fitting of the X-ray pair distribution function of solid solution $Fe_{0.7}Ni_{0.3}B$, $R_w$=0.144. For the refinement, FeB structural model with unit cell parameters obtained upon Rietveld refinement of PXRD data of $Fe_{0.7}Ni_{0.3}B$ was utilized.

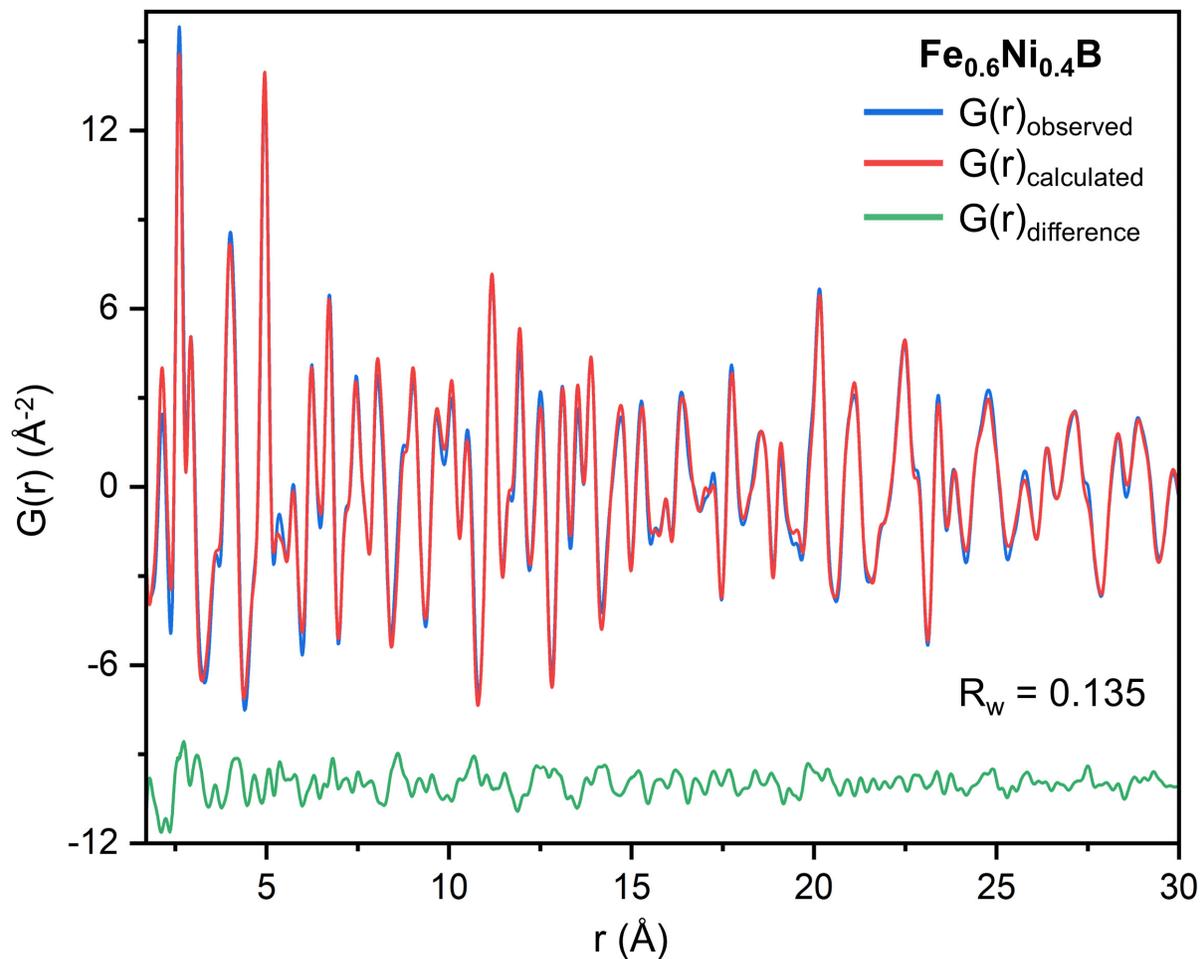

**Figure S4.** Fitting of the X-ray pair distribution function of solid solution $Fe_{0.6}Ni_{0.4}B$, $R_w$=0.135. For the refinement, FeB structural model with unit cell parameters obtained upon Rietveld refinement of PXRD data of $Fe_{0.6}Ni_{0.4}B$ was utilized.

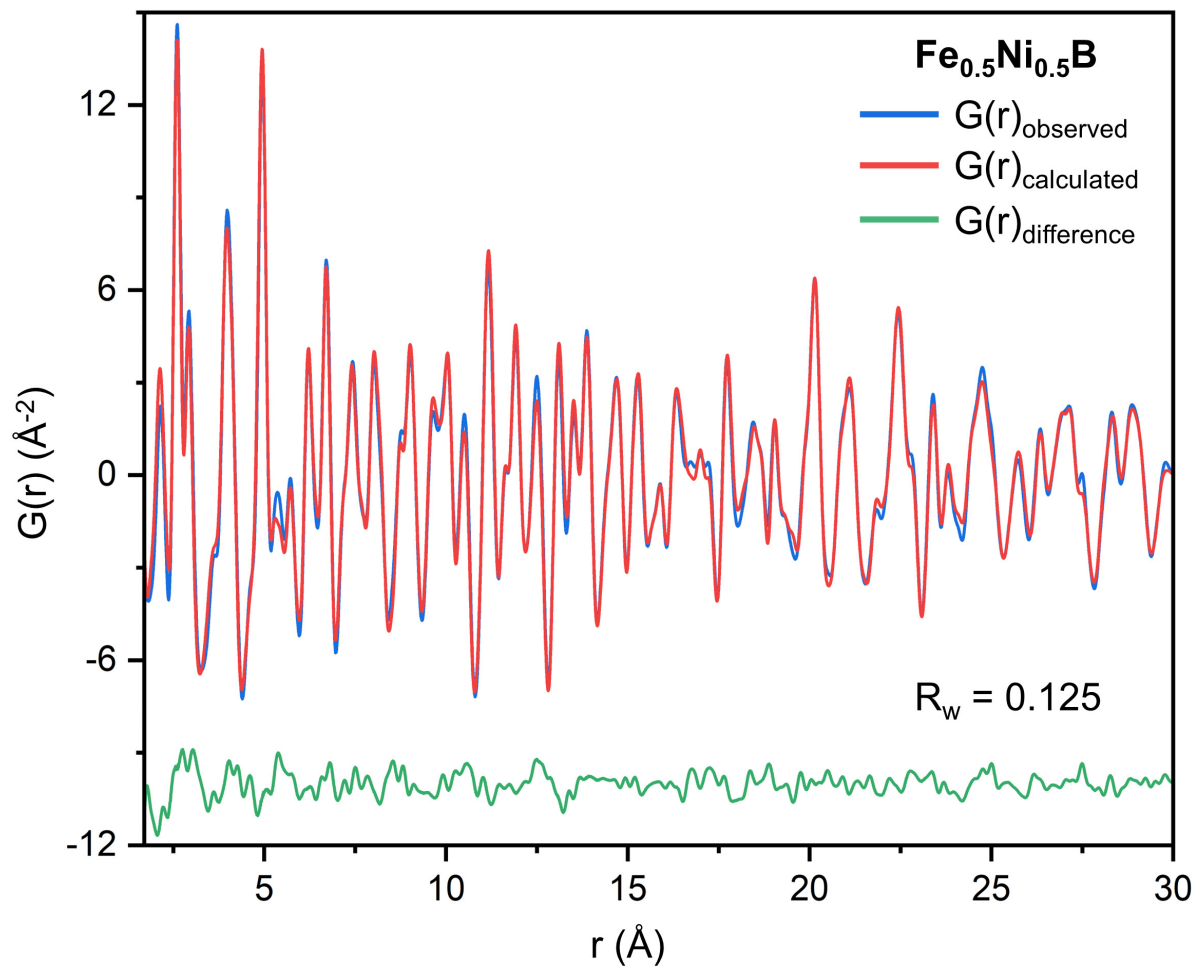

**Figure S5.** Fitting of the X-ray pair distribution function of solid solution $Fe_{0.5}Ni_{0.5}B$, $R_w=0.125$. For the refinement, FeB structural model with unit cell parameters obtained upon Rietveld refinement of PXRD data of $Fe_{0.5}Ni_{0.5}B$ was utilized.

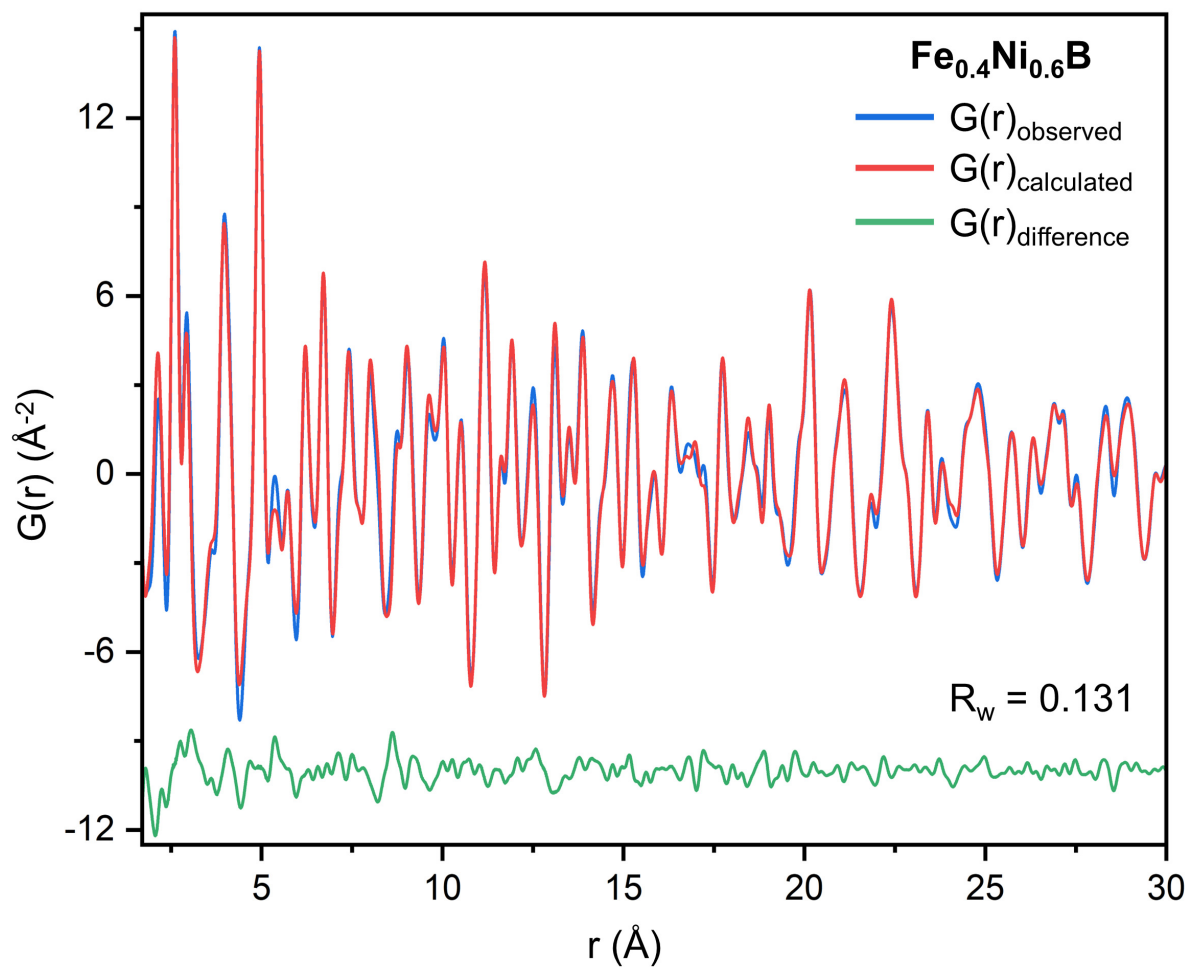

**Figure S6.** Fitting of the X-ray pair distribution function of solid solution $Fe_{0.4}Ni_{0.6}B$, $R_w=0.131$. For the refinement, FeB structural model with unit cell parameters obtained upon Rietveld refinement of PXRD data of $Fe_{0.4}Ni_{0.6}B$ was utilized.

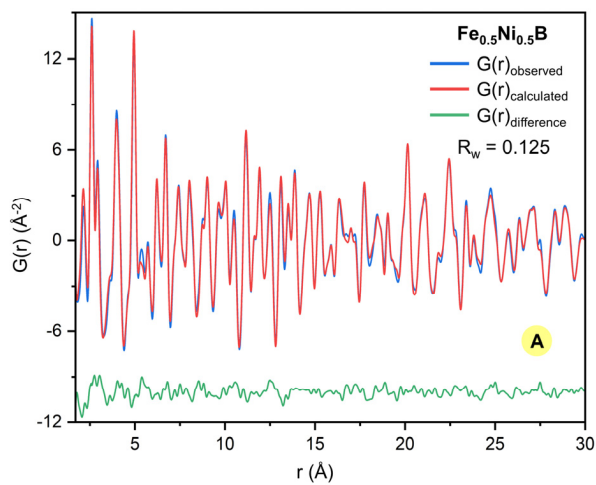
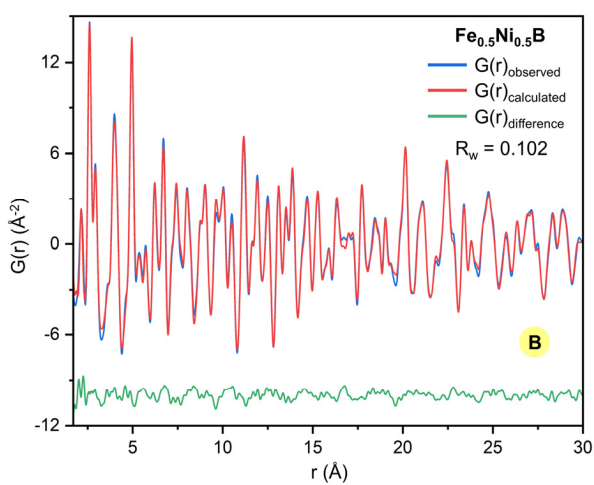
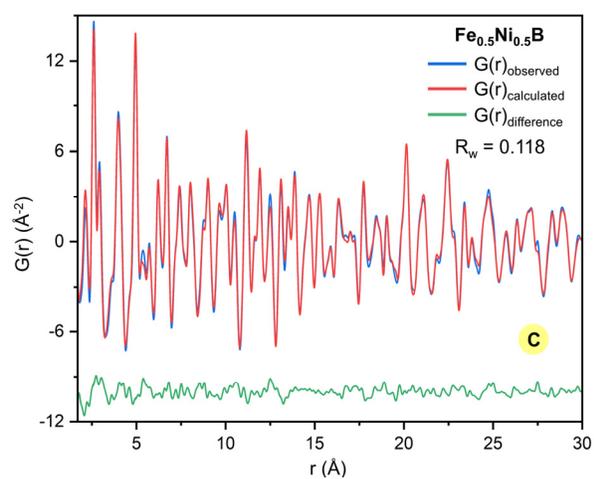

**Figure S7.** Fitting of the X-ray pair distribution function of solid solution Fe$_{0.5}$Ni$_{0.5}$B with

A) FeB structural model (*Pnma*) obtained upon Rietveld refinement of the PXRD data of Fe$_{0.5}$Ni$_{0.5}$B (**no-correlations > 0.8**),

B) FeB structural model (*Pnma*) with unit cell parameters obtained upon Rietveld refinement of PXRD data of Fe$_{0.5}$Ni$_{0.5}$B with a fixed occupancy of Fe:Ni= 0.5:0.5 (**4 correlations > 0.8**),

C) Theoretically predicted structural model of FeNiB$_2$ (*P2$_1$/m*) (**5 correlations > 0.8**).

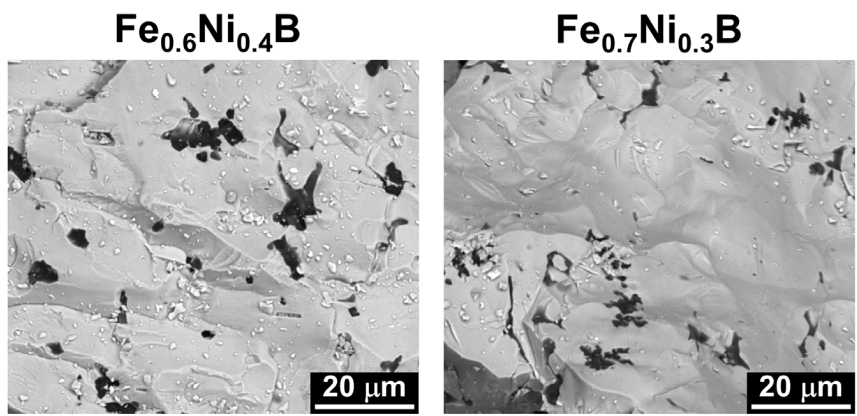

Backscattered Electron (BSE) images

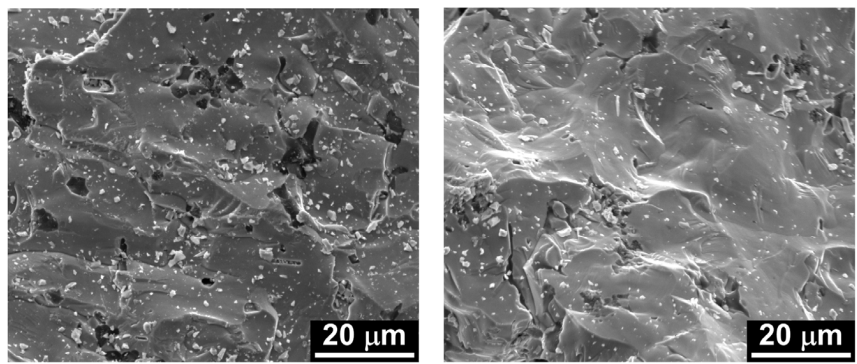

Secondary Electron (SE) images

**Figure S8**. Backscattered (BSE) and secondary electron (SE) images of $Fe_{0.6}Ni_{0.4}B$ and $Fe_{0.7}Ni_{0.3}B$ confirm the presence of fractures in the bulk phase. These fractures appear as dark patches in the BSE images.

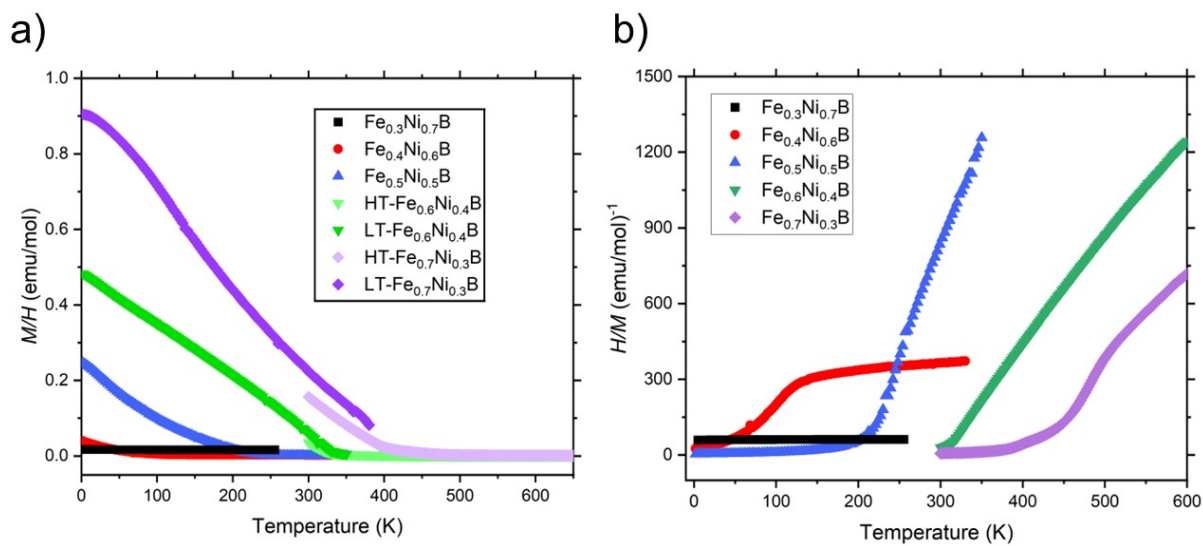

**Figure S9**. (a) Temperature dependence of *M/H* in applied field of 0.1 T for the $Fe_{1-x}Ni_xB$, x = 0.7, 0.6, 0.5, 0.4, 0.3 compounds measured upon cooling. Both high-temperature (HT, measured in PPMS-Oven) and low-temperature (LT, measured in SQUID) data are shown for the Fe-rich samples. (b) *H/M* vs. T for the $Fe_{1-x}Ni_xB$, x = 0.7, 0.6, 0.5, 0.4, 0.3 compounds.

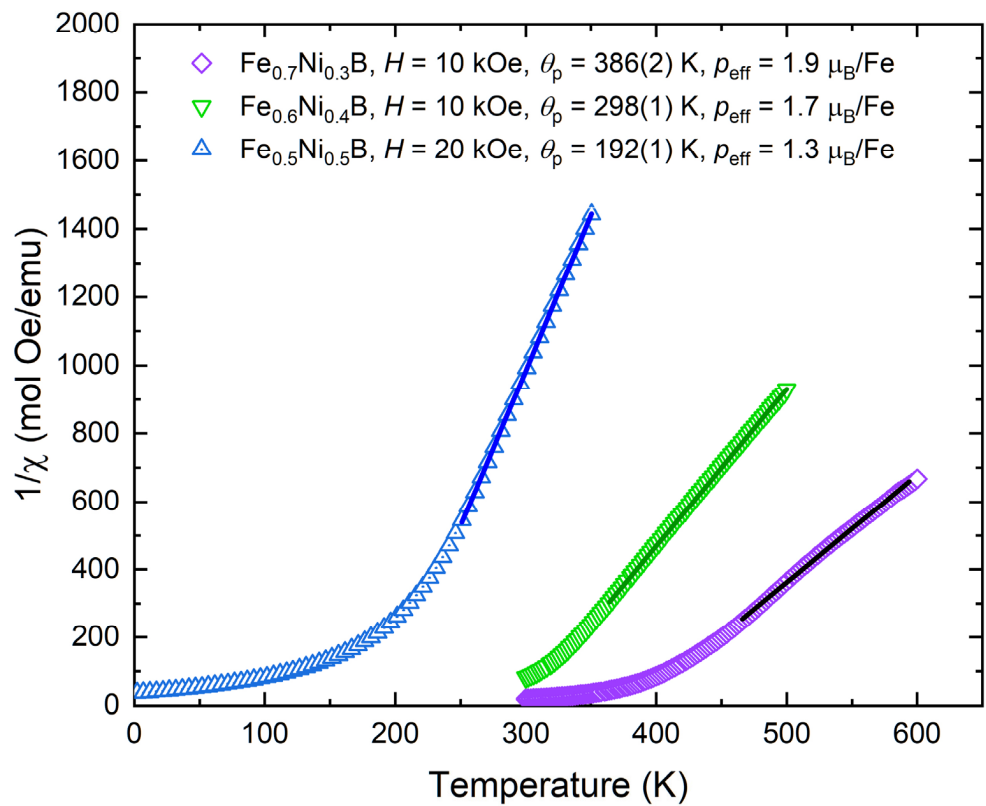

**Figure S10.** Inverse magnetic susceptibility (1/χ vs. *T*) for the Fe$_{1-x}$Ni$_x$B, x = 0.5, 0.4, 0.3 compounds measured in 10 kOe (x=0.3 and 0.4) and 20 kOe (x=0.5) magnetic fields. The Curie-Weiss fits are shown by solid lines.

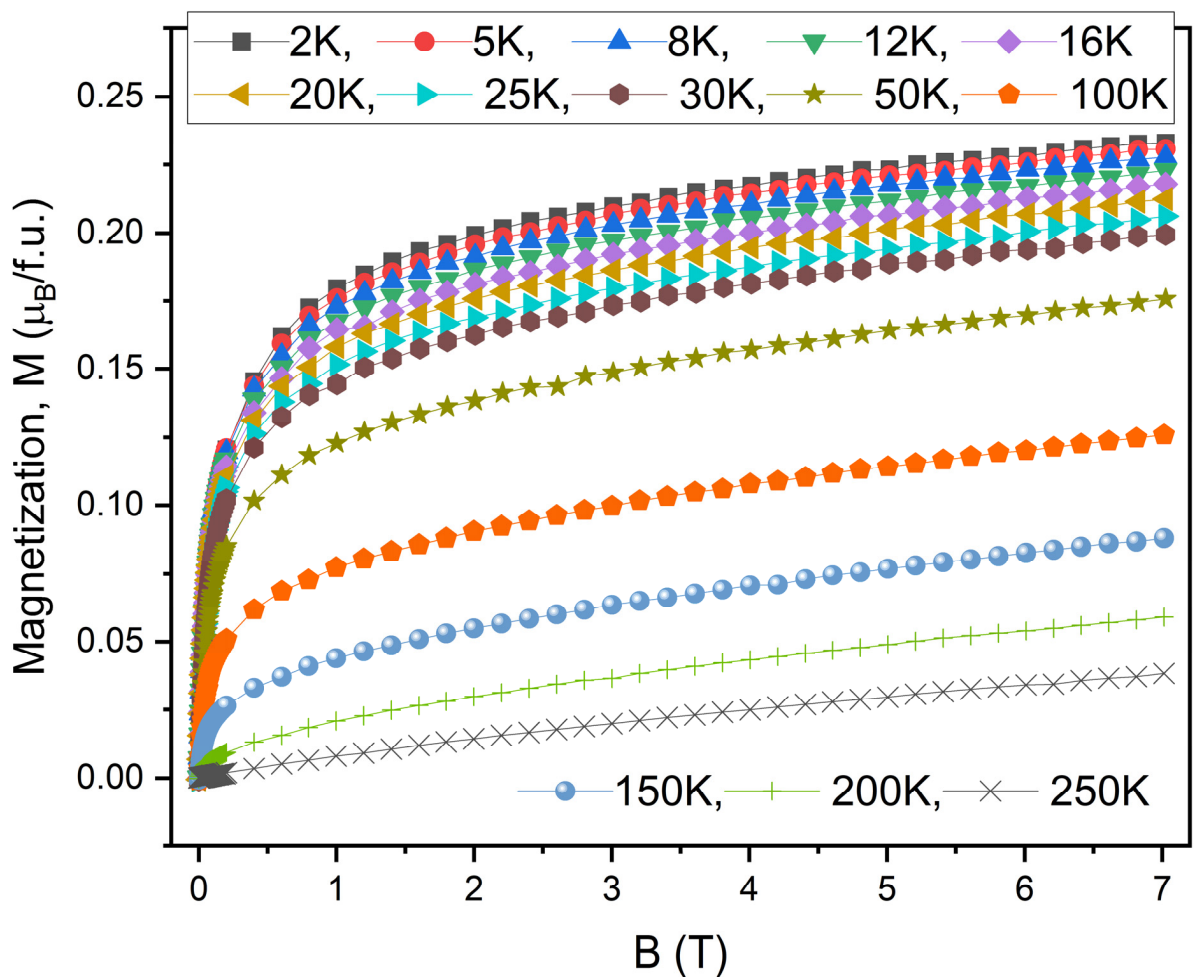

**Figure S11**. Magnetization of Fe$_{0.5}$Ni$_{0.5}$B as a function of applied magnetic field measured at various temperatures.